\documentclass[oldversion]{aa}

\usepackage{graphicx}
\usepackage{graphics}
\usepackage{txfonts}
\usepackage{natbib}
\bibpunct{(}{)}{;}{a}{}{,} % to follow A\&A style

\begin{document}

%macros
\newcommand{\simgt}{\lower.5ex\hbox{$\;\buildrel>\over\sim\;$}}
\newcommand{\simlt}{\lower.5ex\hbox{$\;\buildrel<\over\sim\;$}}
\newcommand{\hst}{{\sl HST}}
\newcommand{\spitzer}{{\sl Spitzer}}
\newcommand{\lya}{{Ly$\alpha$}}                      
\newcommand{\bra}{{Br$\alpha$}}                      
\newcommand{\brg}{{Br$\gamma$}}                      
\newcommand{\ha}{{H$\alpha$}}                      
\newcommand{\hb}{{H$\beta$}}             
\newcommand{\zsun}{{$Z_\odot$}}                      
\newcommand{\msun}{{$M_\odot$}}                      
\newcommand{\lsun}{{$L_\odot$}}                      
\newcommand{\av}{{$A_V$}}
\newcommand{\vi}{{$V-I$}}
\newcommand{\hii}{{H{\sc ii}}}
\newcommand{\hi}{{H{\sc i}}}
\newcommand{\Ne}{{N$_e$}}
\newcommand{\magsq}{mag\,arcsec$^{-2}$}
\newcommand{\micron}{\,$\mu$m}
%\newcommand{\arcsec}{$^{\prime\prime}$}
%\newcommand{\fdg}{\mbox{\ensuremath{.\!\!^\circ}}} % fractional degree symbol: 0.°0

% spectral lines
\newcommand{\nii}{\ensuremath{\mathrm{[N\,II]}}}
\newcommand{\oiii}{\ensuremath{\mathrm{[O\,III]}}}
\newcommand{\oii}{\ensuremath{\mathrm{[O\,II]}}}
\newcommand{\sii}{\ensuremath{\mathrm{[S\,II]}}}

\newcommand{\Ho}{\ensuremath{\mathrm{H}_0}}
\newcommand{\Msun}{\ensuremath{~\mathrm{M}_\odot}}
\newcommand{\Lsun}{\ensuremath{~\mathrm{L}_\odot}}
\newcommand{\LBsun}{\ensuremath{~\mathrm{L}_{B\odot}}}
\newcommand{\AV}{\ensuremath{\mathrm{A}_V}}
\newcommand{\QH}{\ensuremath{Q(\mathrm{H})}}
% BHs
\newcommand{\bh}{\ensuremath{\mathrm{BH}}}
\newcommand{\mbh}{\ensuremath{M_\mathrm{BH}}}
\newcommand{\mdyn}{\ensuremath{M_\mathrm{dyn}}}
\newcommand{\lsph}{\ensuremath{L_\mathrm{bul}}}
\newcommand{\LV}{\ensuremath{L_\mathrm{V}}}
\newcommand{\lopt}{\ensuremath{L_\mathrm{opt}}}

% Units

%
%%
\newcommand{\mlr}{\ensuremath{\Upsilon}}
\newcommand{\I}{\ensuremath{i}}
\newcommand{\Th}{\ensuremath{\theta}}
\newcommand{\B}{\ensuremath{b}}
\newcommand{\So}{\ensuremath{\mathrm{s}_\circ}}
\newcommand{\Vsys}{\ensuremath{V_\mathrm{sys}}}
\newcommand{\chisq}{\ensuremath{\chi^2}}
\newcommand{\chisqr}{\ensuremath{\chi^2_\mathrm{red}}}
\newcommand{\chisqc}{\ensuremath{\chi^2_\mathrm{c}}}
\newcommand{\nnn}{NGC\,5953}

\title{Molecular Gas in NUclei of GAlaxies (NUGA) \\ 
XIII. The interacting Seyfert 2/LINER galaxy NGC\,5953
\thanks{Based on observations
carried out with the IRAM Plateau de Bure Interferometer. IRAM is supported by the
INSU/CNRS (France), MPG (Germany), and IGN (Spain).}
}

\author{V. Casasola \inst{1,2}, 
L.K. Hunt \inst{1}, 
F. Combes \inst{3}, 
S. Garc\'ia-Burillo \inst{4}, 
F. Boone \inst{3}, 
A. Eckart \inst{5}, 
R. Neri \inst{6}, 
\& E. Schinnerer \inst{7}
}

\offprints{{\tt casasola@arcetri.astro.it}}

\institute{
INAF-Osservatorio Astrofisico di Arcetri, Largo E. Fermi 5, 50125 Firenze, Italy
\and INAF-Istituto di Radioastronomia, Via P. Gobetti 101, 40129 Bologna, Italy
\and
Observatoire de Paris, LERMA, 61 Av. de l'Observatoire, F-75014, Paris, France 
\and Observatorio Astron\'omico Nacional (OAN) - Observatorio de Madrid, C/ Alfonso XII, 3, 28014 Madrid, Spain
\and I. Physikalisches Institut, Universit\"at zu K\"oln, Z\"ulpicherstrasse 77, 50937-K\"oln, Germany
\and IRAM-Institut de Radio Astronomie Millim\'etrique, 300 Rue de la Piscine,
38406-St.Mt.d`H\`eres, France
\and Max-Planck-Institut f\"ur Astronomie, K\"onigstuhl 17, D-69117 Heidelberg, Germany
}

\date{Received ; accepted}

\abstract{
We present $^{12}$CO(1--0) and $^{12}$CO(2--1) maps of the 
interacting Seyfert 2/LINER galaxy \nnn\ obtained 
with the IRAM interferometer at resolutions of 2\farcs1 $\times$ 1\farcs4 
and 1\farcs1 $\times$ 0\farcs7, respectively. 
We also present single-dish IRAM 30\,m observations 
of the central region of \nnn\ for the $^{12}$CO(1--0), 
$^{12}$CO(2--1), and HCN(1--0) transitions at resolutions of 22\arcsec,
12\arcsec, and $29^{\prime\prime}$, respectively. 
The CO emission is distributed 
over a disk of diameter $\sim$16\arcsec ($\sim$2.2\,kpc),
within which are several, randomly distributed peaks.
The strongest peak does not coincide with the nucleus,
but is instead offset from the center,
$\sim$2-3$^{\prime\prime}$ ($\sim$340\,pc) toward the west/southwest.
The kinematics of the molecular component are quite regular, as is typical 
of a rotating disk. 
We also compared the $^{12}$CO distribution of \nnn\ with observations 
at other wavelengths in order to study correlations between different 
tracers of the interstellar medium. 
The \hst/F606W WFPC2 images show flocculent spiral structures 
and an ``S-shape'' feature $\simgt$60\,pc in radius, possibly 
associated with a nuclear bar or with the radio jet.
A two-dimensional bulge/disk decomposition of the $H$-band (\hst/F160W) and 
3.6\,$\mu$m (\spitzer/IRAC) images reveals a circumnuclear 
``ring'' $\sim$10-14\arcsec\ in diameter, roughly coincident in size with the 
CO disk and with a star-forming ring previously identified in ionized gas. 
This ring is not present in the near-infrared (NIR) $J-K$ color image, 
nor is it present in the ``dust-only'' image constructed from the 8\,\micron\ IRAC map.
The implication is 
that the excess residual ring is stellar, with colors similar to the surrounding disk. 
We interpret this ring, visible in ionized gas, which appears as stars in the NIR, 
and with no sign of hot dust, as due to a red super giant population 
at least 10--15\,Myr old. However, star formation is still ongoing in the disk
and in the ring itself.
Using NIR images, we computed the gravity torques 
exerted by the stellar potential on the gas. 
The torques are predominantly positive in both  
$^{12}$CO(1--0) and $^{12}$CO(2--1), suggesting that gas 
is not flowing into the center, and less than 5\% of the gas 
angular momentum is exchanged in each rotation. This comes from the 
regular and almost axisymmetric total mass and gas distributions 
in the center of the galaxy.
In \nnn, the AGN is apparently not being actively fueled in the current epoch.
}

\authorrunning{Casasola et al.} 
\titlerunning{NUGA: XIII. NGC\,5953}

\maketitle

\section{Introduction}

Since molecular gas is the predominant phase of the interstellar 
medium (ISM) in the inner regions of spiral galaxies, CO lines represent 
an optimum tracer of nuclear gas dynamics, active galactic nuclei (AGN) 
fueling mechanisms, and their link with circumnuclear star formation.
Although most galaxies host super massive black holes (SMBHs) and the gas accretion 
phenomenon is usually invoked to explain nuclear activity in galaxies, 
the nature of this activity is still not well known.
The feeding of an AGN through accretion depends on an adequate supply of gas
whose angular momentum has been reduced enough to enable
inflow on the small spatial scales surrounding the BH.
Angular momentum must be removed from the disk gas 
\citep[e.g.,][and references therein]{jogee06}, a process that can be 
accomplished through non-axisymmetric perturbations of internal or 
external origin.
In the first case, they arise from disk instabilities and, in the second, 
from galaxy collisions, mergers, and mass accretion \citep{heckman86}.
Either way, they usually manifest themselves as density waves, such as 
large-scale spirals or as bars and their gravity torques \citep[e.g.,][]{sakamoto99,francoise01}, 
or as more localized phenomena, including nested nuclear bars 
\citep[e.g.,][]{friedli93}, lopsidedness or $m = 1$ 
perturbations \citep[e.g.,][]{shu90,santi00}, or warped nuclear 
disks \citep[e.g.,][]{schinnerer00a,schinnerer00b}.

To better understand  the mechanisms for gas fueling of AGN, 
we started a high-resolution and high-sensitivity CO survey
of nearby active galaxies at the IRAM Plateau de Bure 
Interferometer (PdBI), the NUclei of GAlaxies (NUGA) project 
\citep[][]{santi03}. The galaxies of the NUGA sample already 
studied show surprising results. 
In fact, there is no unique circumnuclear molecular gas feature 
linked with nuclear activity, but instead a variety of molecular gas morphologies 
which characterize the inner kpc of active galaxies.
These morphologies include one- and two-armed instabilities \citep{santi03},
well-ordered rings and nuclear spirals \citep{francoise04,vivi08}, 
circumnuclear asymmetries \citep{melanie05} and large-scale bars 
\citep{fred07,leslie08}.
The analysis of the torques exerted by the stellar gravitational potential 
on the molecular gas of the NUGA sample has shown that the gas can be driven 
away from the AGN (e.g. for NGC\,4321) or toward it 
(e.g. for NGC\,2782, NGC\,3147, and NGC\,4579).
However, the velocities observed for NUGA are
too small to correspond to the AGN feedback models 
where violent molecular outflows and superwinds are expected
\citep[e.g][]{narayanan06,hopkins06}.

The different morphologies we find are probably related to the various 
timescales \citep{santi05}. Large-scale bars can transport gas inward
efficiently \citep[e.g.,][]{francoise85,sakamoto99}, 
and it appears that they can also drive powerful starbursts 
\citep[e.g.,][]{knapen02,jogee05}.
Nevertheless, a clear correlation between large-scale bars and nuclear 
activity has not yet been
found \citep[e.g.,][]{mulchaey97}, probably because the timescales for bar-induced
gas inflow and AGN duty cycles are very different.
Bars drive inflow over timescales \citep[$\gtrsim$300 Myr,][]{jogee05} larger than
those of AGN accretion-rate duty cycles \citep[$\sim$1-10 Myr,][]{heckman04,hopkins06,king07}, 
and active accretion seems to occur only intermittently over the lifetime of a
galaxy \citep[][]{ferrarese01,marecki03,janiuk04,hopkins06,king07}.
This implies that most AGN are in an intermediate phase between active 
accretion episodes, making the detection of galaxies with nuclear accretion 
rather difficult.

Viscosity rather than self-gravity can also play a significant role 
in the fueling process.
Viscous torques, generally weak and with timescales quite long at large radii,
in combination with gravitational torques can coordinate efforts to produce
recurrent episodes of activity during the typical lifetime of any
galaxy \citep{santi05}.
Viscous torques can produce gas inflow on scales $\sim$100-200 pc 
if they act on a contrasted nuclear ring distribution and in the absence of particularly 
strong positive gravitational torques.
In NGC\,4579, the efficiency of viscosity may be comparable
to the efficiency of gravity torques in the 
inner $\sim$50\,pc \citep[][]{santi09}.

This paper, dedicated to the galaxy \nnn, is the latest of the NUGA 
series where results obtained for the galaxies of the sample are described 
on a case-by-case basis.
\nnn\ ($D$ = 28 Mpc for $H_{0}$ = 73 km s$^{-1}$Mpc$^{-1}$) is an interacting 
galaxy \citep[e.g.,][]{rampazzo95,vivi04,iono05}, 
classified as a Seyfert 2 by \citet{gonzalez96} and as a LINER 
by \citet{veilleux95}, and of early and unbarred Hubble type 
(SAa pec).
\nnn\ and its late-type companion NGC\,5954 
(LINER/Seyfert 2, SAB(rs)cd pec) constitute a binary system (VV\,244, Arp\,91) 
where the two galaxies are separated by a projected distance of 
5.8 kpc ($\sim$43\arcsec).
They show clear signs of interaction visible in the 
distorted morphology, the presence of a tidal bridge (or distorted arm) 
connecting the two galaxies, and of prominent star-forming regions.
Both galaxies have circumnuclear starbursts that may have been 
induced by the interaction \citep{gonzalez96}.

\nnn\ hosts a compact radio core and jet, revealed by high-resolution
radio continuum observations with MERLIN \citep{melanie07a}.
The jet is resolved at 18\,cm,
and after beam deconvolution is roughly $\sim$0\farcs3 in length (40\,pc), 
with a position angle (PA) of $\sim$10$^\circ$.
The small-scale radio continuum structure at 20\,cm is similar in orientation, but 
with lower spatial resolution ($\sim$1\farcs5), slightly 
more toward the east, PA$\sim$25$^\circ$ \citep{jenkins84}.
This jet-like elongation as seen at lower resolution
was first referred to as a ``jet'' by
\citet{gonzalez96} because it is roughly aligned with the structure
in the ``excitation map'', obtained by dividing [O{\sc iii}] emission
by H$\alpha$. 

The NGC\,5953/54 pair has also been mapped in atomic and molecular gas.
Both galaxies are embedded in a common \hi\ envelope,
with a clear velocity gradient along the \hi\ plume extending
more than 8\,kpc to the northwest \citep{chengalur94,iono05,haan07,haan08}.
There is some indication of a faint diffuse optical counterpart
to the \hi\ plume \citep{chengalur94,hernandez03}.
The \hi\ peak is also significantly displaced from 
the stellar disks \citep{iono05}.
The overall \hi\ velocity gradient runs from southeast to northwest, 
roughly perpendicular to the rotation in the ionized and
molecular gas \citep{hernandez03,iono05}.
The most recent \hi\ mass determination for \nnn\ has been
obtained by  \citet[][]{haan08}, M$_{\rm H\,I}=1.1\times$10$^{9}$\,M$_{\odot}$ 
(value scaled to our adopted distance of $D=28$\,Mpc), typical of that 
expected for interacting galaxies of the same morphological type 
\citep[][]{vivi04}.

\begin{table}
\caption[]{Fundamental parameters for \nnn.}
\begin{center}
\begin{tabular}{lll}
\hline
\hline
Parameter  & Value$^{\mathrm{b}}$ & References$^{\mathrm{c}}$ \\
\hline
$\alpha_{\rm J2000}$$^{\mathrm{a}}$ & 15$^h$34$^m$32.36$^s$ & (1) \\
$\delta_{\rm J2000}$$^{\mathrm{a}}$ & 15$^{\circ}$11$^{\prime}$37\farcs70 & (1) \\
$\alpha_{\rm dyn}$$^{\mathrm{a}}$   & 15$^h$34$^m$32.38$^s$ & (2) \\
$\delta_{\rm dyn}$$^{\mathrm{a}}$   & 15$^{\circ}$11$^{\prime}$37\farcs59 & (2) \\
$V_{\rm sys, hel}$              & 1990 km\,s$^{-1}$ & (1) \\
RC3 Type                        & SAa pec & (3) \\
Nuclear Activity                & S2/LINER & (4) $\&$ (5) \\
Inclination                     & 42$^{\circ}$ & (1) \\
Position Angle                  & 45$^{\circ}$ $\pm$ 1$^{\circ}$\ & (1) \\
Distance                        & 28\,Mpc ($1^{\prime\prime} = 136\,{\rm pc}$) & (3) \\
L$_{B}$                         & $4.9 \times 10^{9}$\,L$_{\odot}$ & (6) \\
M$_{\rm H\,I}$                  & $1.1 \times 10^{9}$\,M$_{\odot}$ & (7) \\
M$_{\rm H_{2}}$                 & $2.2 \times 10^{9}$\,M$_{\odot}$ & (8) \\
M$_{\rm dust}$(60 and 100\,$\mu$m)& $2.7 \times 10^{6}$\,M$_{\odot}$ & (6) \\
L$_{\rm FIR}$                   & $1.4 \times 10^{10}$\,L$_{\odot}$ & (9) \\
\hline
\hline
\end{tabular}
\label{table1}
\end{center}
\begin{list}{}{}
\item[$^{\mathrm{a}}$] ($\alpha_{\rm J2000}$, $\delta_{\rm J2000}$) is the
phase tracking center of our $^{12}$CO observations, 
($\alpha_{\rm dyn}$, $\delta_{\rm dyn}$) 
is the dynamical center derived from radio observations for the core 
of \nnn\ by \citet{melanie07a}.
\item[$^{\mathrm{b}}$] 
Luminosity and mass values extracted from the literature
have been scaled to the distance of $D=28$\,Mpc.
\item[$^{\mathrm{c}}$] (1) This paper;
(2) \citet{melanie07a};
(3) NASA/IPAC Extragalactic Database (NED); 
(4) \citet{gonzalez96}; (5) \citet{veilleux95}; 
(6) \citet{vivi04}; (7) \citet{haan08}; (8) \citet{iono05}; 
(9) {\it IRAS} Catalog.
\end{list} 
\end{table}

The molecular gas distribution in \nnn\ is symmetric with a 
short extension pointing toward NGC\,5954 \citep[e.g.,][]{yao03,iono05}.
The H$_{2}$ mass content estimated by \citet{iono05} is 
%$1.6 \times 10^{9}\,M_{\odot}$ , 
$2.2\times10^{9}\,$M$_{\odot}$ (scaled to the distance of $D=28$\,Mpc
for \nnn), higher than the molecular hydrogen mass
found for the interacting companion NGC\,5954
%($M_{\rm H_{2}}$ = $7.7 \times 10^{8}\,M_{\odot}$).
(M$_{\rm H_{2}}$ = $1.1 \times 10^{9}\,$M$_{\odot}$).
Higher-order transitions of the CO molecule have also been detected in \nnn, 
including the $^{12}$CO(3--2) line by \citet{yao03} with the James Clerk 
Maxwell Telescope (FWHM$\sim$15\arcsec), suggesting a high excitation 
of the carbon monoxide, indicative of dense and hot gas.
Table \ref{table1} summarizes the fundamental characteristics of
\nnn.

The structure of this paper is as follows.  In Sect. \ref{sec:obs}, 
we describe our new observations of \nnn\ and the literature data 
with which we compare them.
In Sects. \ref{sec:30m} and \ref{sec:pdbi}, we present the observational 
results, both single dish and interferometric, describing morphology,
excitation conditions, and kinematics of the molecular gas in the inner kpc of
\nnn.
Comparisons between $^{12}$CO observations and those obtained at other
wavelengths are given in Sect. \ref{sec:stars}.
In Sect. \ref{sec:torques}, we describe the computation
of the gravity torques derived from the stellar potential 
in the inner region of \nnn.
%Sect. \ref{sec:discussion} discusses some results presented
%in previous sections, which are summarized
%in Sect. \ref{sec:conclusions}.
Sect. \ref{sec:conclusions} summarizes our results.

We will assume a distance to \nnn\ of $D=28$\,Mpc
\citep[HyperLeda\footnote{http://leda.univ-lyon1.fr},][]{paturel03}
and a Hubble constant $H_0=73$\,km\,s$^{-1}$\,Mpc$^{-1}$.
This distance implies that 1\arcsec\ corresponds to 136\,pc.
 
\begin{figure*}
\centering
\includegraphics[width=0.8\textwidth,angle=-90]{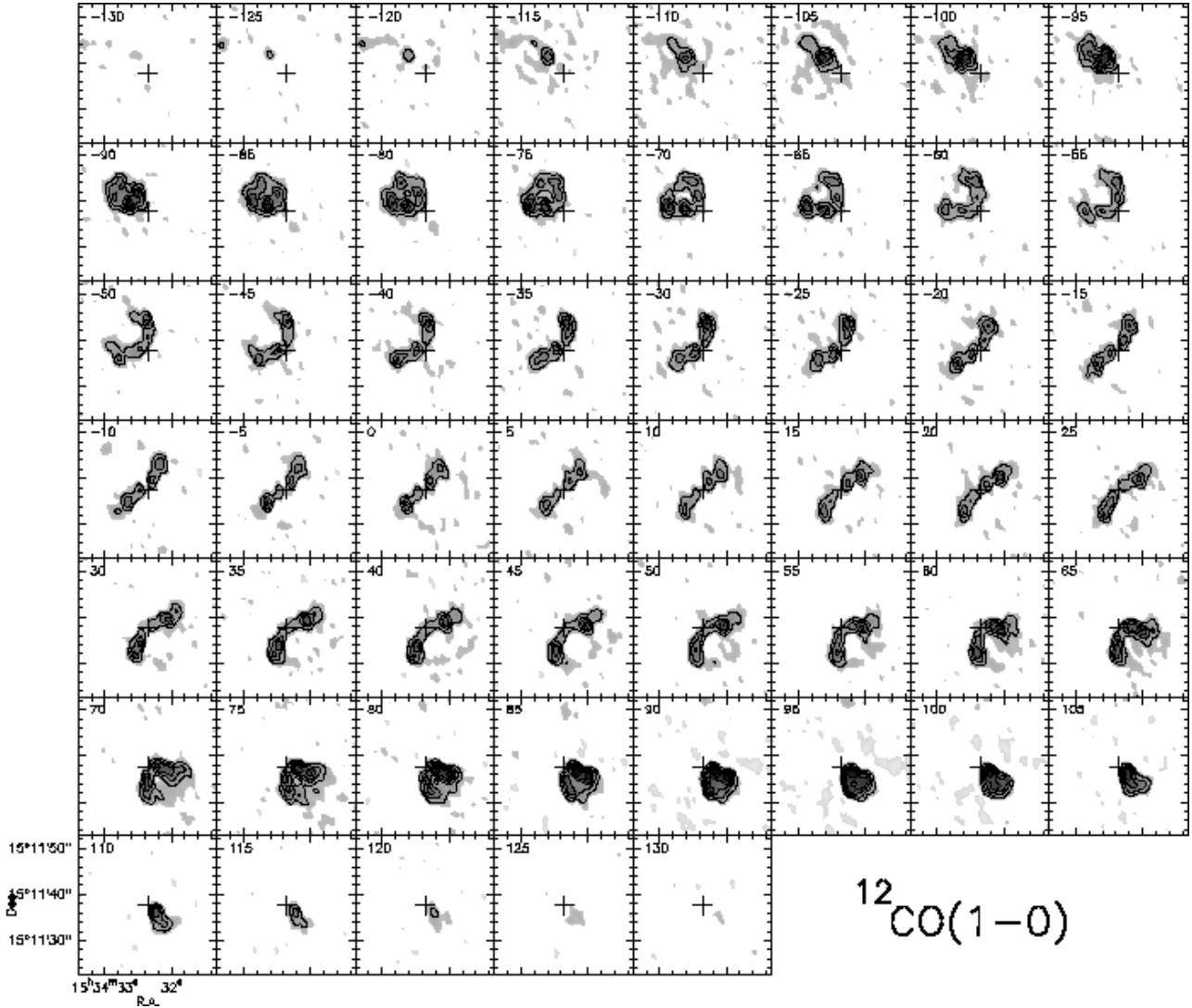}
\caption{$^{12}$CO(1--0) velocity channel maps observed with the 
IRAM PdBI+30\,m in the nucleus of \nnn, with a spatial resolution 
of 2\farcs1 $\times$ 1\farcs4 (HPBW).
The maps are centered on the phase tracking center of our
observations assumed to be coincident with the dynamical center of the galaxy
(see Sect. \ref{sec:dyncen}).
Velocity channels range from $\Delta V = -130\,{\rm km\,s^{-1}}$ 
to $+130\,{\rm km\,s^{-1}}$ in steps of $5\,{\rm km\,s^{-1}}$ relative 
to $V_{\rm sys, hel}$ = 1990 km\,s$^{-1}$ (see Sect. \ref{sec:dyncen}).
The contours run from 
$-0.20\,{\rm mJy\,beam^{-1}}$ to $120\,{\rm mJy\,beam^{-1}}$ 
with spacings of $20\,{\rm mJy\,beam^{-1}}$.}
\label{channels10}
\end{figure*}

\begin{figure*}
\centering

\includegraphics[width=0.8\textwidth,angle=-90]{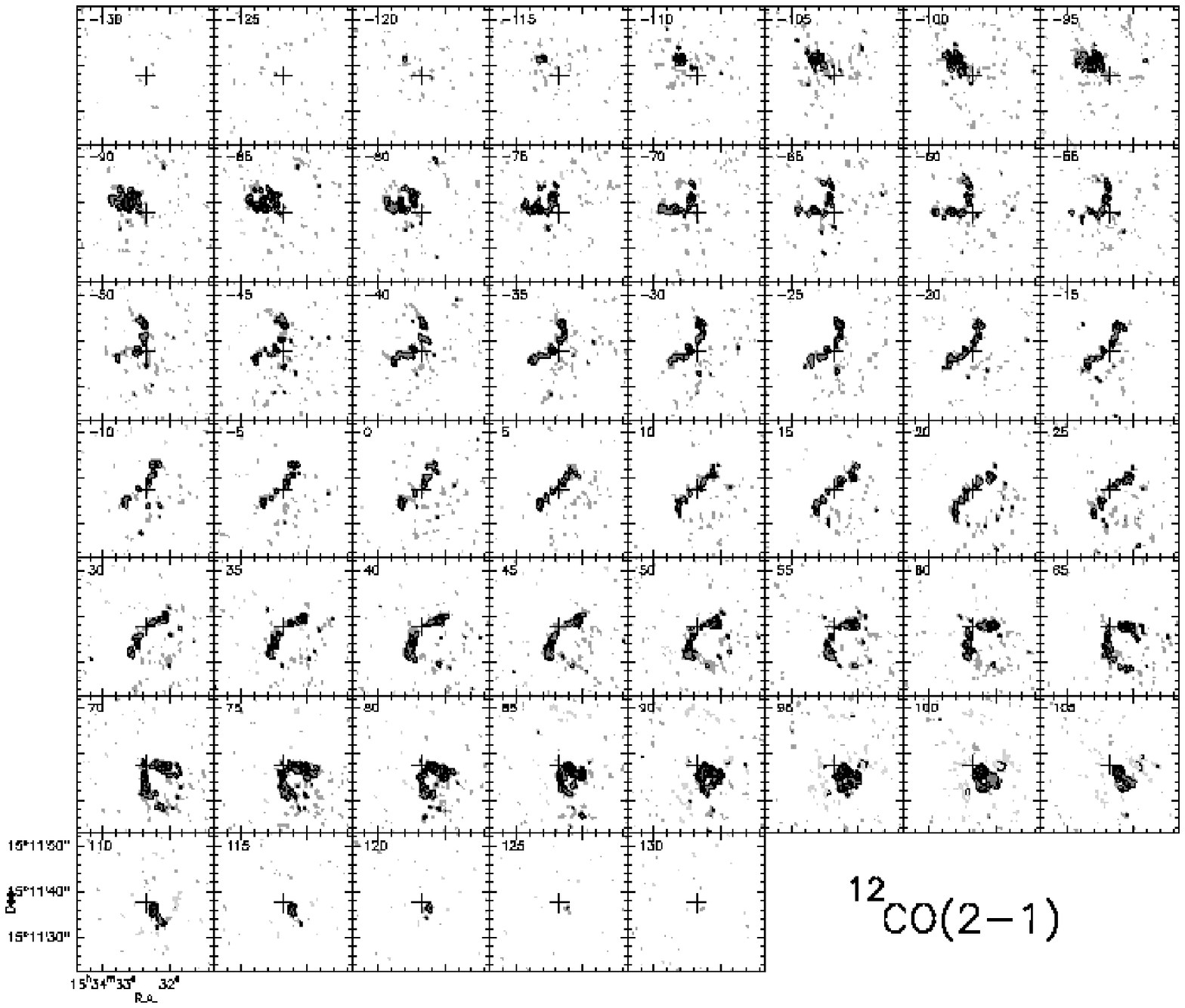}
\caption{Same as Fig. \ref{channels10} but for the $^{12}$CO(2--1) line, 
with a spatial resolution of 1\farcs1 $\times$ 0\farcs7.
The contours run from $-60\,{\rm mJy\,beam^{-1}}$ to 
$150\,{\rm mJy\,beam^{-1}}$ with spacings of $30\,{\rm mJy\,beam^{-1}}$.}
\label{channels21}
\end{figure*}

\section{Observations \label{sec:obs}}
\subsection{Interferometric observations \label{sec:PdB-obs}}

We observed \nnn\ with the IRAM PdBI in the ABCD configuration 
of the array between September 2003 and February 2004 in the 
$^{12}$CO(1--0) (115\,GHz) and the $^{12}$CO(2--1) (230\,GHz)
rotational transitions.
The PdBI receiver characteristics, the observing procedures, and the 
image reconstruction are the same as described in \citet{santi03}.
The quasars 3C454.3 and  3C273 were used for bandpass and flux 
calibrations respectively, and the quasar 1546+027 was used for 
phase and amplitude calibrations.

Data cubes with 512 $\times$ 512 pixels (0\,\farcs20\,${\rm pixel}^{-1}$ 
for $^{12}$CO(1--0) and 0\,\farcs14\,${\rm pixel}^{-1}$ for $^{12}$CO(2--1)) 
were created over a velocity interval of -242 km\,s$^{-1}$ to 
+242 km\,s$^{-1}$ in bins of 5 km\,s$^{-1}$. 
The images were reconstructed using the standard 
IRAM/GILDAS\footnote{http://www.iram.fr/IRAMFR/GILDAS/} 
software \citep{guilloteau} and restored with gaussian beams of dimensions 
2\farcs0 $\times$ 1\farcs4 (PA = $25^{\circ}$) at 115\,GHz 
and 1\farcs1 $\times$ 0\farcs7 (PA = $22^{\circ}$) at 230\,GHz.
We used natural weighting to generate the $^{12}$CO(1--0) maps
and uniform weighting to generate $^{12}$CO(2--1) maps.
Such a procedure maximizes the flux recovered in $^{12}$CO(1--0)
and optimizes the spatial resolution in $^{12}$CO(2--1).
In the cleaned maps, the {\it rms} uncertainty $\sigma$
in 5\,km\,s$^{-1}$ width velocity channels is 2.7\,mJy\,beam$^{-1}$
and $6.0\,{\rm mJy\,beam^{-1}}$ for the $^{12}$CO(1--0) and
$^{12}$CO(2--1) lines, respectively.
At a level of $\sim$3$\sigma$
no 3\,mm (1\,mm) continuum was detected toward \nnn, down to an {\it rms}
noise level of 0.32\,mJy\,beam$^{-1}$ (0.48\,mJy\,beam$^{-1}$). 
The conversion factors between intensity and brightness temperature 
are $32\,{\rm K\,(Jy\,beam^{-1})^{-1}}$ at 115\,GHz and 
$28\,{\rm K\,(Jy\,beam^{-1})^{-1}}$ at 230\,GHz.
All velocities are referred to the systemic velocity
$V_{\rm sys, hel}$ = 1990 km\,s$^{-1}$ 
(see later Sect. \ref{sec:dyncen}) and 
$(\Delta \alpha, \Delta \delta)$ offsets are relative to 
the phase tracking center of the observations 
($\alpha_{J2000}$, $\delta_{J2000}$) = (15$^h$34$^m$32.36$^s$, 
15$^{\circ}$11$^{\prime}$37\farcs70). 
All maps presented in this paper are centered on this position.
The maps are not corrected for primary beam attenuation.

\subsection{Single dish observations and short spacing correction \label{sec:30m-obs}}

We performed IRAM 30\,m telescope observations in a $5 \times 5$
raster pattern with 7\arcsec\ spacing from 16 to 19 
July 2002.
We used 4 SIS receivers to observe simultaneously 
at the frequencies of the $^{12}$CO(1--0) (115\,GHz), 
the $^{12}$CO(2--1) (230\,GHz), and the HCN(1--0) (89\,GHz) lines.
The half power beam widths are 22\arcsec\ for $^{12}$CO(1--0),
12\arcsec\ for $^{12}$CO(2--1), and 29\arcsec\ for
HCN(1--0).
The CO(2--1) line has been observed in dual-polarization.
Typical system temperatures were $\sim$270-390\,K at 115\,GHz,
$\sim$300-750\,K at 230\,GHz, and $\sim$120\,K at 89\,GHz.
Throughout this paper the line intensity scale is expressed in units of 
$T_{\rm mb}$, the beam-averaged radiation temperature. 
$T_{\rm mb}$ is related to $T^{*}_{A}$, the equivalent antenna temperature 
reported above the atmosphere, by $\eta=T^{*}_{\rm A}/T_{\rm mb}$ where
$\eta$ is the telescope main-beam efficiency.
At 115\,GHz $\eta$ = 0.79, at 230\,GHz $\eta$ = 0.54, and at 89\,GHz 
$\eta$ = 0.82.
All observations were performed in ``wobbler-switching'' mode, with a minimum
phase time for spectral line observations of 2\,s and a maximum beam throw 
of $240^{\prime\prime}$. The pointing accuracy was 
$\sim$3$^{\prime\prime}$ {\it rms}.
The single dish maps are centered on the phase tracking center
of the interferometric observations (see Table \ref{table1}).

Single dish $^{12}$CO observations were used to compute short spacings 
and complete the interferometric measurements.
We combined 30\,m and PdBI data using the SHORT-SPACE task available in 
the GILDAS software.
The best compromise between good angular resolution and complete
restoration of the missing extended flux was found by varying the
relative weights  of 30\,m and PdBI observations.
These were chosen in order to obtain the same mean weights in the 
single dish data as in the interferometric data within a ring in 
the $uv$ plane ranging from $1.25\,D/\lambda$ to $2.5\,D/\lambda$
($D = 15$\,m).
The combined  PdBI+30\,m data have produced maps with angular
resolutions of 2\farcs1 $\times$ 1\farcs4 at PA = $25^{\circ}$ 
for the $^{12}$CO(1--0) and 1\farcs1 $\times$ 0\farcs7 at 
PA = $22^{\circ}$ for the $^{12}$CO(2--1).
In the combined maps, the {\it rms} uncertainty $\sigma$
in 5\,km\,s$^{-1}$ width velocity channels is 2.5\,mJy\,beam$^{-1}$
and 5.5\,mJy\,beam$^{-1}$ for the $^{12}$CO(1--0) and
$^{12}$CO(2--1) lines, respectively.
All figures presented in this paper are made with 
short-spacing-corrected data.

We estimate that the $^{12}$CO(1--0) map including 
only PdBI observations within $\sim$22\arcsec\ 
($^{12}$CO(1--0) HPBW for the 30\,m telescope) recovers  
a flux of S${\rm _{CO(1-0)}}$ = 150 Jy km\,s$^{-1}$, 
$\sim$81$\%$ of the total flux 
measured with the combined PdBI+30\,m map, S${\rm _{CO(1-0)}}$ = 185 Jy km\,s$^{-1}$. 
Table \ref{table2} collects $^{12}$CO(1--0) flux values, both 
present in literature and determined with our observations, 
for \nnn.
In this table, Col. (1) indicates the reference, Cols. (2) and (3) 
the telescope, both single dish and interferometer, and the diameter 
of the single dish telescope, Col. (4) is the primary beam of the instrument
or the diameter used for the performed photometry, Col. (5) is the beam
in interferometric measurements, and Col. (6) is the measured flux.
Table \ref{table2} shows that $^{12}$CO(1--0) fluxes we obtained with interferometric 
observations, single dish, and combined measurements (PdBI+30\,m)
are in good mutual agreement with each other and with data present in literature.
Our $^{12}$CO(1--0) combined map within 42\arcsec\ gives a value (254 Jy\,km\,s$^{-1}$) 
consistent with the flux of 233 Jy km\,s$^{-1}$ found with OVRO \citep{iono05}, 
but consistent also with the NRAO flux of 365  Jy\,km\,s$^{-1}$ \citep{zhu99} 
taking into account that they used a single dish with a diameter of 12\,m and 
a primary beam of 55\arcsec.
In addition, we recovered $\sim$79$\%$ of the total flux detected by \citet{young95} 
with the FCRAO (320 Jy\,km\,s$^{-1}$), a reasonable agreement considering the 
uncertainties in the amplitude calibration and the non-correction by the 
primary beam attenuation.

\begin{table*}
\caption[]{$^{12}$CO(1--0) flux values for \nnn.}
\begin{center}
\begin{tabular}{llllll}
\hline
\hline
Reference & Telescope &	Diameter & Primary beam	or FOV$^{\mathrm{a}}$ & Beam & Flux \\
& & [m] &	[\arcsec] & [\arcsec $\times$ \arcsec] &	[Jy km\,s$^{-1}$] \\
\hline
\citet{iono05}	& OVRO     &    & 60 & 4.4 $\times$ 3.6 & 233	\\
\citet{zhu99}	& NRAO     & 12 & 55 &                  & 365	\\
\citet{young95}	& FCRAO	   & 14 & 45 &		        & 320	\\
This paper	& PdBI+30\,m &	& 42 & 2.1 $\times$ 1.4	& 254	\\
This paper	& PdBI+30\,m & 	& 22$^{\mathrm{b}}$ & 2.1 $\times$ 1.4	& 185	\\
This paper	& PdBI	   &	& 22$^{\mathrm{b}}$ & 2.0 $\times$ 1.4	& 150	\\
This paper	& 30\,m	   & 30	& 22 (central position) & & 167$^{\mathrm{c}}$	\\
This paper	& 30\,m	   & 30	& 22 (inner $50^{\prime\prime} \times 50^{\prime\prime}$) & &377$^{\mathrm{d}}$ \\
\hline
\hline
\end{tabular}
\label{table2}
\end{center}
\begin{list}{}{}
\item[$^{\mathrm{a}}$] 
Primary beam is considered for single dish observations, while field-of-view (FOV) 
for interferometric or combined (interferometric+single dish) ones.
\item[$^{\mathrm{b}}$] 
The photometry has been performed within 22\arcsec, the $^{12}$CO(1--0) primary 
beam for the 30\,m telescope.
\item[$^{\mathrm{c}}$] 
The $^{12}$CO(1--0) recovered flux for the central position (0\arcsec, 0\arcsec).
\item[$^{\mathrm{d}}$] 
The $^{12}$CO(1--0) recovered flux for inner 
$50^{\prime\prime} \times 50^{\prime\prime}$, $5 \times 5$ mapping with 7\arcsec spacing (see Sect. \ref{sec:30m-obs}).
\end{list} 
\end{table*}

Figures \ref{channels10} and \ref{channels21} show the channel maps 
for the $^{12}$CO(1--0) and $^{12}$CO(2--1) lines, respectively,
in the central region of \nnn.
Figures \ref{fig:co21-30m} and \ref{fig:n5953-hcn} display the single 
dish data, for the $^{12}$CO(1--0), $^{12}$CO(2--1), and HCN(1--0) 
lines.
The two $^{12}$CO lines have been mapped on a $5 \times 5$ grid
with $7^{\prime\prime}$ spacings, while the HCN(1--0) line has been mapped
on a $3 \times 3$ grid with $7^{\prime\prime}$ spacings and the nine
HCN(1--0) spectra have been averaged to improve the signal-to-noise.

\subsection{Optical and infrared images \label{sec:otherdata}}

We first acquired a broad-band optical image from the \hst\ 
archive\footnote{Based on observations made with the NASA/ESA 
Hubble Space Telescope, and obtained from the Hubble Legacy Archive,
which is a collaboration between the Space Telescope Science 
Institute (STScI/NASA), the Space Telescope European Coordinating 
Facility (ST-ECF/ESA) and the Canadian Astronomy Data Centre 
(CADC/NRC/CSA).} of \nnn\ obtained with the F606W filter 
(mean wavelength of 5940\,$\AA$).
This image 
was first published by \citet{malkan98}, in
a survey of 256 of the nearest (z$\leq$0.035) Seyfert 1, 
Seyfert 2, and starburst galaxies.
The image 
covers the inner $\sim$$20^{\prime\prime}\times20^{\prime\prime}$ 
with a pixel size of 0\farcs045.

From the \hst\ archive, we  also acquired the \hst\ NICMOS F160W ($H$-band, 
$1.6\,{\rm \mu m}$) image of \nnn. 
This near-infrared (NIR) image 
was presented by \citet{regan99} and \citet{leslie04b}, and
%The image has been obtained with an exposure time of  640\,s and
covers the inner $\sim$$19^{\prime\prime}\times19^{\prime\prime}$
of the galaxy with a pixel size of 0\farcs075.

We also acquired infrared (IR) images obtained with the IRAC camera
on \spitzer, available thanks to the project
`Starburst Activity in Nearby Galaxies'
(Principal Investigator: G. Rieke).
The IRAC images, from 3.6, to $8\,{\rm \mu m}$, were reduced 
with MOPEX \citep{makovoz05} which accounts for distortion and rotates 
to a fiducial coordinate system. 
They cover a large sky area ($\sim$320$^{\prime\prime} \times 320^{\prime\prime}$) 
including both \nnn\ and NGC\,5954.
We imposed a pixel size of 1\farcs20 for the final images, 
roughly the same as the original IRAC detector.
Following \citet{helou04}, a ``dust-only'' (non-stellar) image was derived 
from the 8\,\micron\ image, with subtraction of the
stellar component computed by scaling the 3.6\,\micron\ and 4.5\,\micron\ images
and subtracting them. 
This image will be referred to as a dust-only image, and should be
dominated by emission from Polycyclic Aromatic Hydrocarbons (PAHs) and perhaps
some hot dust continuum emission.

A $J-K$ image was derived from NIR data published 
by \citet{leslie99b} acquired with
ARNICA (Arcetri Near-Infrared Camera) mounted on the
Nordic Optical Telescope (NOT\footnote{The NOT is operated on 
the island of La Palma jointly by Denmark, Finland, Norway, 
and Sweden, in the Spanish Observatorio del Roque de los Muchachos
of the Instituto de Astrofisica de Canarias.}).
ARNICA was an imaging camera for the NIR bands between
1.0 and 2.5 $\mu$m based on the Rockewell HgTeCd-array
detector NICMOS 3 (256$\times$ 256 pixels), and
designed and built by Arcetri Observatory (Firenze, Italy)
for the Infrared Telescope at Gornergrat in Switzerland.
%(TIRGO, 1.5\,m diameter).
% Later, ARNICA was mounted on other telescopes, such as
% the William Herschel Telescope at La Palma (WHT, 4.2\,m) and
% the NOT telescope.
The ARNICA/NOT NIR images ($J$, $H$, $K$) cover 
$\sim$120$^{\prime\prime}\times120^{\prime\prime}$,
including the pair of interacting galaxies, 
NGC\,5953 and NGC\,5954, with a pixel size of 0\farcs546.

\begin{figure*}
\centering
\hbox{
\includegraphics[width=0.5\textwidth,angle=0]{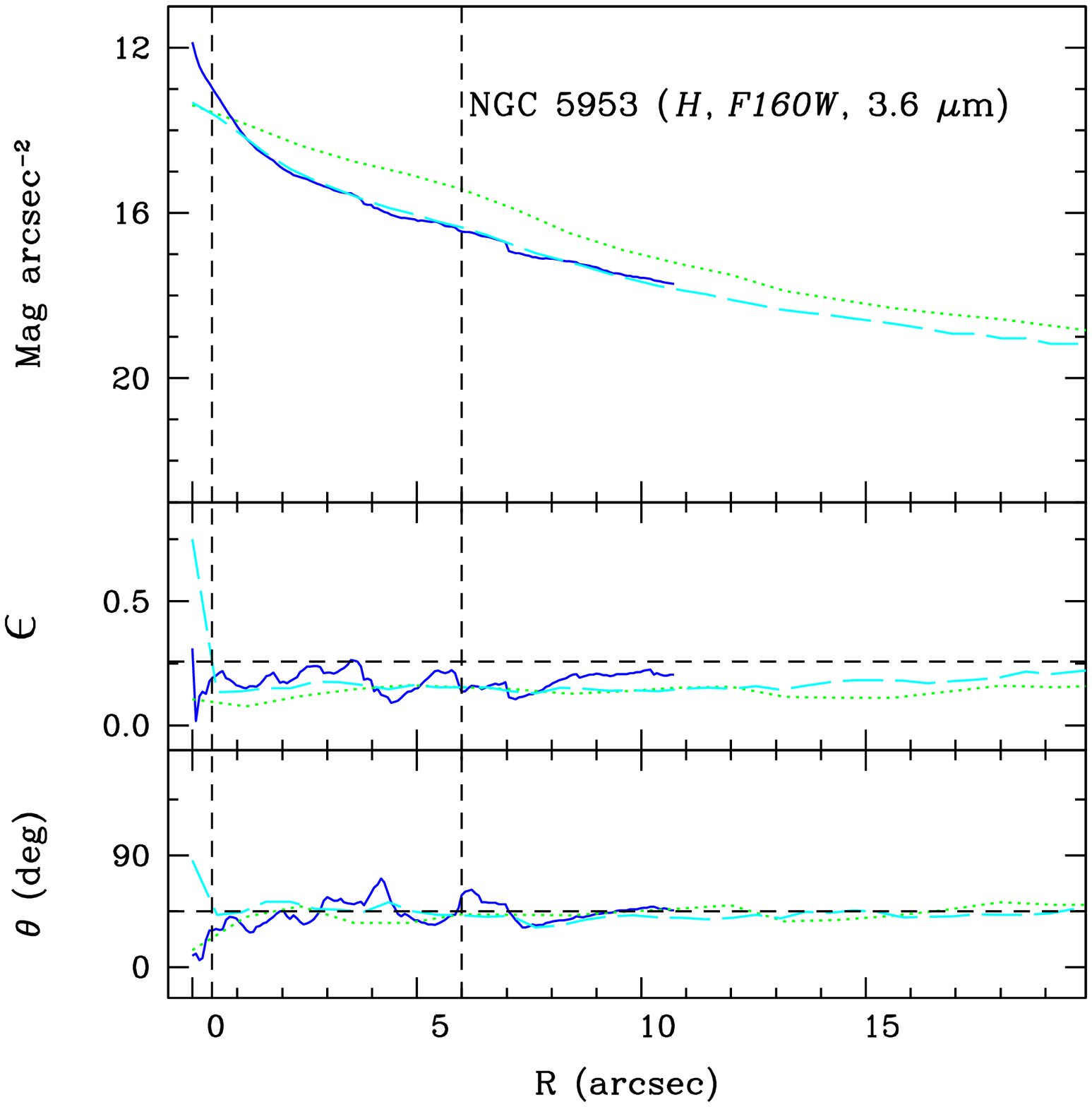}
\includegraphics[width=0.5\textwidth,angle=0]{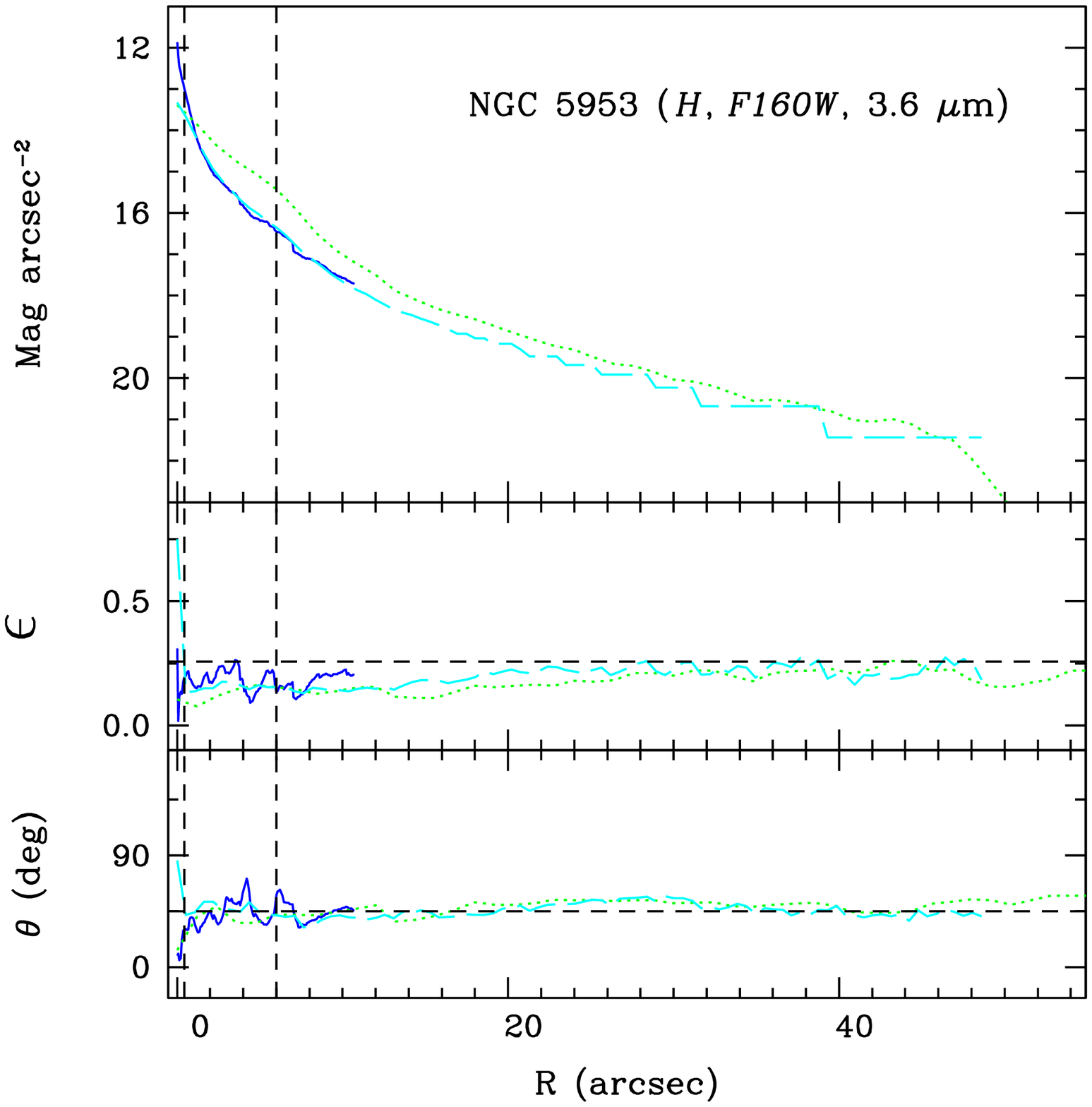}
}
\caption{Radial surface brightness profiles of \nnn\ made by elliptical
averages.
The dotted line corresponds to IRAC 3.6\,\micron, the dashed line to
ground-based $H$-band, and the solid line to \hst/NICMOS/F160W.
The left panel shows a smaller FOV than the right panel.
The adopted position angle (45$^\circ$) 
and the best-fit inclination (42$^\circ$) are shown by dashed horizontal
lines in the lower panels (see Sect. \ref{sec:dyncen}).
The dashed vertical lines correspond to the 60\,pc ($\sim$0\farcs44, radius) nuclear bar seen in the \hst\ F606W image (see Sect. \ref{sec:othermorphology}), and the 820\,pc (radius) ring observed
in the residuals of the bulge-disk decomposition (see Sect. \ref{sec:stellar}). 
\label{fig:profiles}
}
\end{figure*}

We compared the surface brightness profiles of the ground-
and space-based images by extracting elliptically averaged profiles,
centered on the brightness peaks.
The position angle and ellipticity were allowed to vary in
the ellipse fitting. 
These radial profiles are shown in Figure \ref{fig:profiles};
the dashed horizontal lines in the lower panels correspond to the
adopted position angle and inclination (see Sect. \ref{sec:dyncen}).
The profiles agree quite well; in particular the ground-based $H$-band 
and \hst\ F160W show the same trend down to the nuclear
regions where the ground-based atmospheric beam smearing leads to
lower surface brightness.
The $H$-band/3.6\,\micron\ color is relatively constant throughout
the entire radial range shown, with slightly redder colors from
$\sim$5-8$^{\prime\prime}$. 

\section{Single dish results\label{sec:30m}}

\begin{figure*}
\centering
\begin{tabular}{c}
\includegraphics[height=0.7\textwidth,angle=-90]{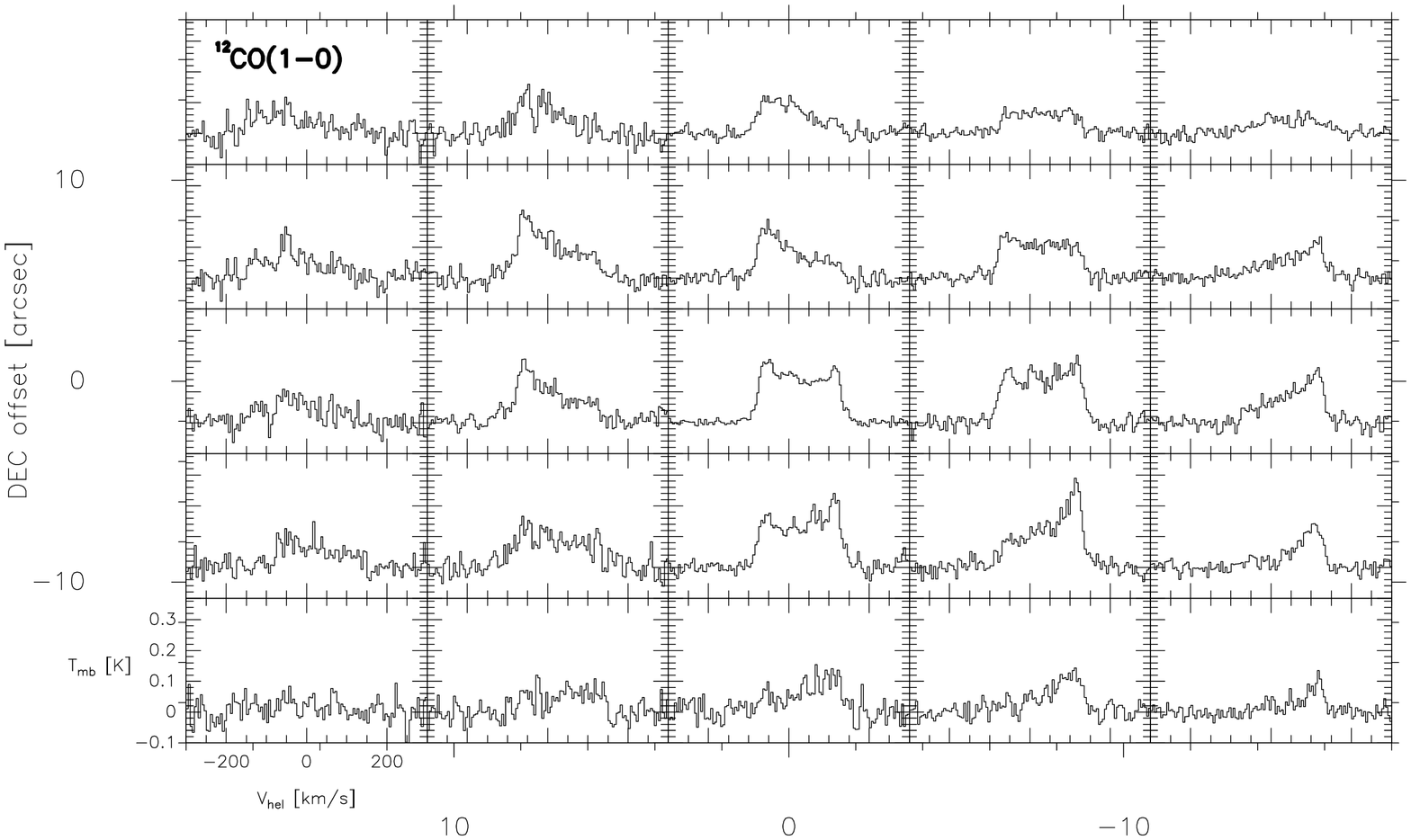}\\
\includegraphics[height=0.7\textwidth,angle=-90]{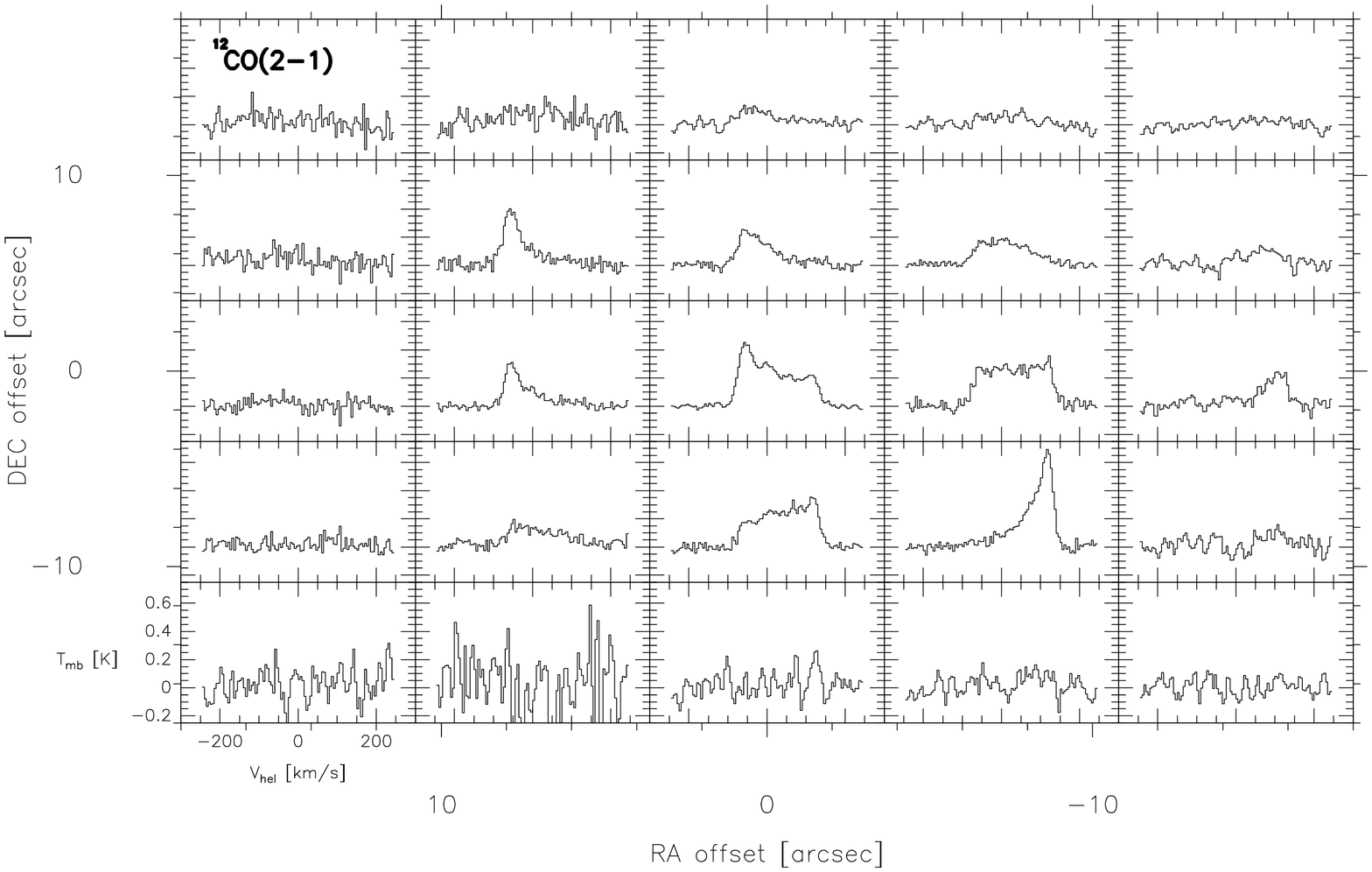}\\
\end{tabular}
\caption{
Spectra maps of \nnn\ made with the IRAM 30\,m with
$7^{\prime\prime}$ spacing in $^{12}$CO(1--0) (top) and $^{12}$CO(2--1) (bottom). 
The positions are arcsec offsets relative to the phase tracking center of our
observations.
Each spectrum has a velocity scale from
$-300$ to $300\,{\rm km\,s^{-1}}$, and a beam-averaged radiation temperature
scale ($T_{\rm mb}$) from $-0.10$ to $0.37\,{\rm K}$ for $^{12}$CO(1--0) and 
from $-0.25$ to 0.75\,K for $^{12}$CO(2--1).
}
\label{fig:co21-30m}
\end{figure*}

The observations performed with the A and B receivers of the IRAM 30\,m 
telescope in the two $^{12}$CO lines covered the
inner $\sim$50$^{\prime\prime}$, corresponding to the central $\sim$6.8 kpc
(in diameter) of the galaxy (Fig. \ref{fig:co21-30m}). 
The 25 observed positions show that the central region of \nnn\ hosts 
extended molecular emission both in $^{12}$CO(1--0) and $^{12}$CO(2--1) emission 
(Fig. \ref{fig:co21-30m}). The maximum detected $T_{mb}$ is 0.18 K in 
$^{12}$CO(1--0) in the southwest corresponding to offset position 
(-7$^{\prime\prime}$,-7$^{\prime\prime}$), and 0.54 K in $^{12}$CO(2--1) 
at the same offset position (-7$^{\prime\prime}$,-7$^{\prime\prime}$).

Within the inner $\sim$50$^{\prime\prime} \times 50^{\prime\prime}$
we estimate a flux of 377 Jy\,km\,s$^{-1}$, in good agreement with previous 
single dish flux determinations \citep[][see Table \ref{table2}]{zhu99,young95}.
From this $^{12}$CO(1--0) integrated flux assuming a H$_{2}$-CO 
conversion factor $X = N(\rm{H_{2}})$/$I\rm_{CO} = 2.2 \times 10^{20}$ cm$^{-2}$ 
(K km s$^{-1}$)$^{-1}$ \citep{solomon91}, we can derive the H$_{2}$ 
mass within the observed region as:
\begin{eqnarray}M\rm_{H_{2}}\mbox{[M$_{\odot}$]} &=& 8.653 \times 10^{3}\,D^{2}\mbox{[Mpc]}\,S\rm_{CO(1-0)}\mbox{[Jy\,km\,s$^{-1}$]}
\label{h2mass}
\end{eqnarray}

\noindent
We obtain M$\rm_{H_{2}}$$\sim$2.6$\times 10^{9} \rm{M_{\odot}}$,
and including the mass of helium, the corresponding total molecular 
mass is M$\rm_{mol} = M\rm_{H_{2}+He} = 1.36 \times M\rm_{H_{2}}$$
\sim$3.5$\times 10^{9} \rm{M_{\odot}}$.

\begin{figure}
\centering
\includegraphics[width=0.29\textwidth,angle=-90]{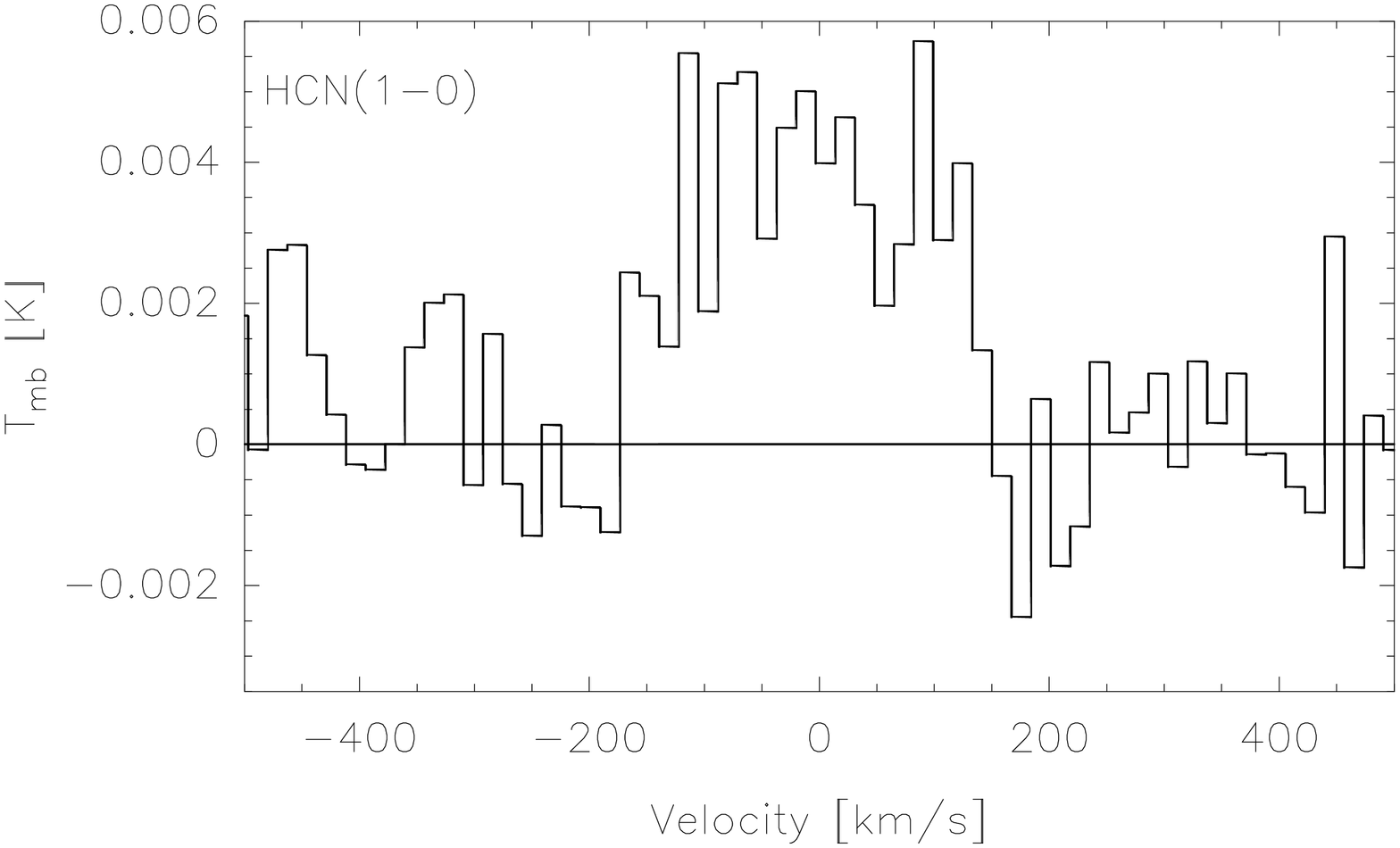}
\caption{
HCN(1--0) spectrum toward the center of \nnn, averaged over
the 9-point map made with the IRAM 30\,m with $7^{\prime\prime}$ spacing.
The spectrum has a velocity scale from $-500$ to $500\,{\rm km\,s^{-1}}$ 
and a beam-averaged radiation temperature scale ($T_{\rm mb}$) from 
$-0.0035$ to $0.006\,{\rm K}$.}
\label{fig:n5953-hcn}
\end{figure}

The HCN(1--0) line has been observed for 9 positions with 7$^{\prime\prime}$ spacing,
covering the central $\sim$43\arcsec\ ($\sim$5.8\,kpc).
The HCN(1--0) emission is detected in the west part of the observed region, and  
the average spectrum over the 3$\times$3 grid is shown in Fig. \ref{fig:n5953-hcn}.
The HCN(1--0) peak is at $T_{\rm mb}$$\sim$0.005\,K. 
The CO(1--0)/HCN(1--0) ratio is very high, equal to 20  on average over the center,
where instead we would expect an enhanced HCN emission due to the AGN 
and thus a lower ratio.
A high CO/HCN line ratio is rarely encountered in AGN but not completely unusual.
NGC\,6951 has a ratio of 30 in the starburst ring and 2.5 in the nucleus 
\citep{melanie07b}, and NGC\,3147 of 20 in the inner 4 kpc
where the CO emission exhibits two ring-like structures around the 
nucleus \citep{vivi08}.
On the other hand, the galaxy NGC\,1097 has a CO/HCN ratio which 
ranges from 3 in the nucleus to 10 in the star-forming ring \citep{kohno03}.
Since AGN activity implies an enhanced HCN emission relative to CO
emission associated with the star-formation process,
the high CO(1--0)/HCN(1--0) ratio
observed in \nnn\ implies
that excitation by star-formation is dominant over AGN excitation 
in the circumnuclear region.

\section{Interferometric results\label{sec:pdbi}}

\subsection{Dynamical center and inclination\label{sec:dyncen}}

We assume that the dynamical center of \nnn\
coincides with the position of the AGN ``core'' derived from 
radio observations (MERLIN/18\,cm) by \citet{melanie07a}:
$\alpha$=15$^h$34$^m$32.38$^s$ and 
$\delta$=15$^{\circ}$11$^{\prime}$37\farcs59.
Since these coordinates are nearly coincident with those of
the phase tracking center of our $^{12}$CO observations 
(see Table \ref{table1}), in the following we assume that our 
observations are centered on the dynamical center of the galaxy.

\begin{figure*}
\centering
\hbox{
\includegraphics[width=0.3\textwidth,angle=-90]{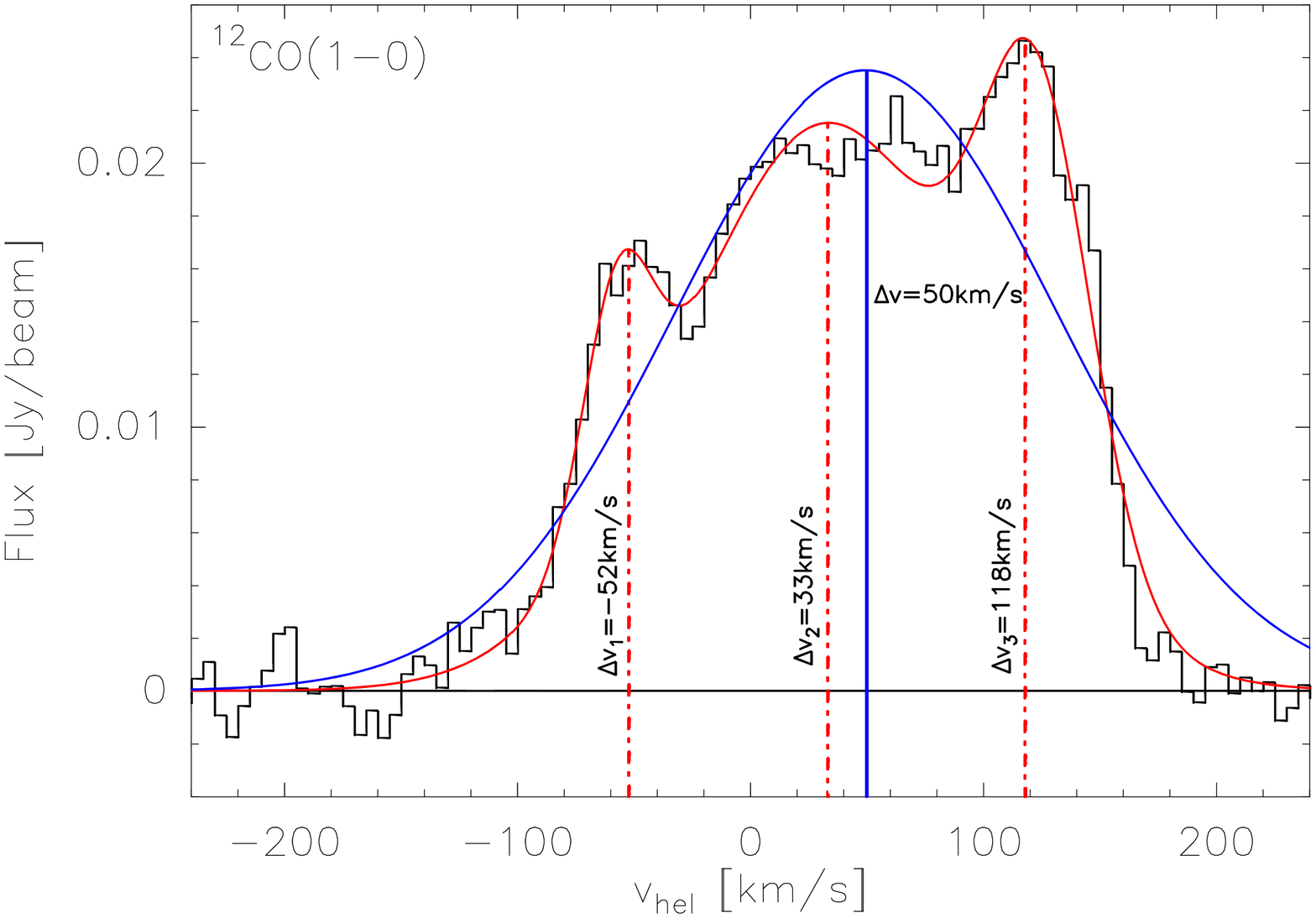}
\hspace{0.1\textwidth}
\includegraphics[width=0.3\textwidth,angle=-90]{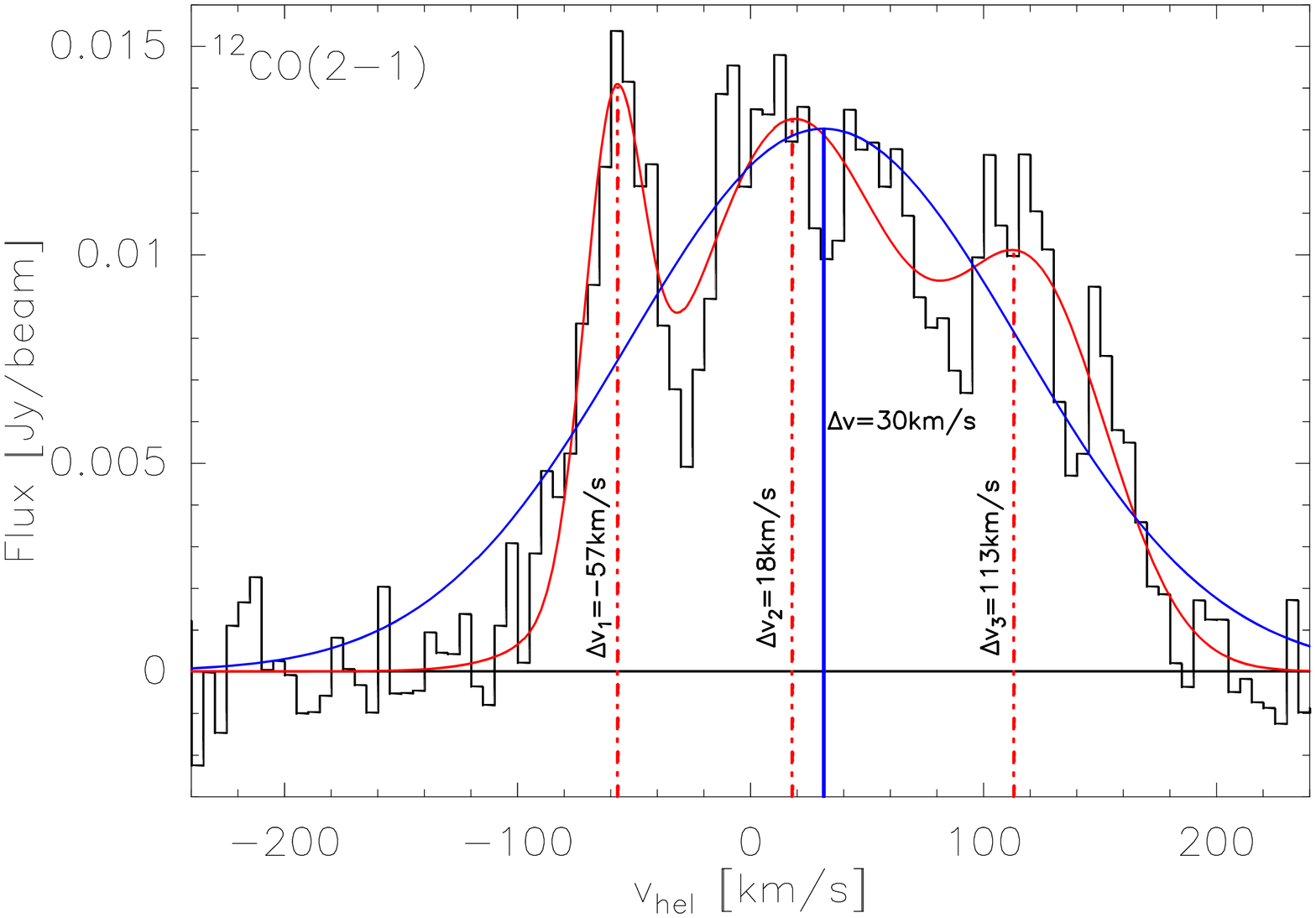}
}
\caption{
\textit{Left panel}: $^{12}$CO(1--0) integrated spectrum, one component 
gaussian (blue) fit, and three components gaussian (red) fit in the inner 
$\sim$4$^{\prime\prime}$ associated with the AGN source for 
PdBI+30\,m combined data. For the one component gaussian fit the 
heliocentric systematic velocity is redshifted by 50 km\,s$^{-1}$ with 
respect to the heliocentric velocity of the center (0 km\,s$^{-1}$).
\textit{Right panel}: Same for $^{12}$CO(2--1). For the one component 
gaussian fit the heliocentric systematic velocity is redshifted by 
30 km\,s$^{-1}$.
\label{fig:velhel}
}
\end{figure*}

The spectral correlators were centered at 114.520 GHz and 
229.037 GHz for the $^{12}$CO(1--0) and $^{12}$CO(2--1) 
line respectively, corresponding to 
$V_{\rm LSR}$ = 1966 km\,s$^{-1}$.
The difference between LSR and heliocentric velocity is
16 km\,s$^{-1}$, and therefore the observations were
centered on $V_{\rm hel}\rm (PdBI)$ = 1950 km\,s$^{-1}$.
In the inner $\sim$4\arcsec\ the velocity centroid  
is 50 km\,s$^{-1}$ redshifted with respect to the heliocentric 
velocity of the center of our $^{12}$CO(1--0) observations 
(Fig. \ref{fig:velhel}, left panel, one component gaussian fit), 
and 30 km\,s$^{-1}$ redshifted relative to $^{12}$CO(2--1) 
(Fig. \ref{fig:velhel}, right panel, one component gaussian fit). 
We therefore estimate the systemic heliocentric velocity 
as the intermediate value between that determined for 
the $^{12}$CO(1--0) and that for the $^{12}$CO(2--1), 
$V_{\rm sys, hel}$ = 1990 km\,s$^{-1}$.

In Figs. \ref{fig:velhel}, we fitted the emissions also considering
three gaussian components, visible both in  $^{12}$CO(1--0) and 
in $^{12}$CO(2--1). These (red) fits show that the wing components 
for the two $^{12}$CO lines are approximately at the same velocities within the 
noise (-52 km s$^{-1}$ for the $^{12}$CO(1--0) vs. -57 km s$^{-1}$ 
for the $^{12}$CO(2--1), and 118 km s$^{-1}$ for the $^{12}$CO(1--0) vs. 
113 km s$^{-1}$ for the $^{12}$CO(2--1)). 
The major difference is present for the central gaussian component, 
33 km s$^{-1}$ for the $^{12}$CO(1--0) vs. 18 km s$^{-1}$ for the 
$^{12}$CO(2--1), maybe due to the clumpy nature of the inner molecular gas.
The line ratio $R_{21}$ ($I_{CO(2-1)}/I_{CO(1-0)}$) assumes the values 
of 1.2, 0.6, and 0.5 respectively for the three components passing 
from negative to positive velocities. 
The mean ratio of 0.8 is consistent with the line ratio discussed 
later in Sect. \ref{sec:coratios}.

The systemic heliocentric velocity of 1990 km\,s$^{-1}$
is 25 km\,s$^{-1}$ redshifted with respect 
to the value determined from \hi\ observations 
(1965 km\,s$^{-1}$, HyperLeda),
and in agreement with the shift of 50 km\,s$^{-1}$
inferred from previous $^{12}$CO(1--0) observations
with OVRO \citep{iono05}.
This discrepancy between systemic velocity derived
from $^{12}$CO and \hi\ observations is not unusual,
especially in galaxies with a lopsided \hi\ morphology.
The NUGA galaxy NGC\,4579 exhibits a difference
of $\sim$50 km\,s$^{-1}$, probably due to the interaction
history of the galaxy and the different effect of the ram-pressure 
on the atomic and molecular gas \citep[][]{santi09}.
Since the atomic gas is much more affected by ram-pressure 
\citep[][]{Kenney86,Vollmer01} than the molecular component,   
the \hi\ kinematics are more sensitive than the CO to
the interaction history;
\nnn\ could present a similar scenario to NGC\,4579.
In fact, in \nnn\ the \hi\ has been detected mainly in the outer 
parts where a big tidal tail connecting the two galaxies shifts the  
\hi\ barycenter with respect to the molecular gas that traces the less perturbed,
inner parts of the galaxy.

We also derived the inclination of \nnn\ by fitting the 
two-dimensional bulge/disk (B/D) decomposition on 
the ground-based $H$-band image and the 
IRAC 3.6\,\micron\ image (see Sect. \ref{sec:stellar}). 
The best-fit inclinations are $i$ = 39$^{\circ}$
and 46$^{\circ}$, respectively;
we therefore used a roughly intermediate value of 42$^{\circ}$,
which approximates quite well the outer regions of the NIR images
(see Fig. \ref{fig:profiles}).

\subsection{CO morphology and mass\label{sec:comorphology}}

\begin{figure*}
\centering
\includegraphics[width=0.43\textwidth,angle=-90,bb=145 17 460 750]{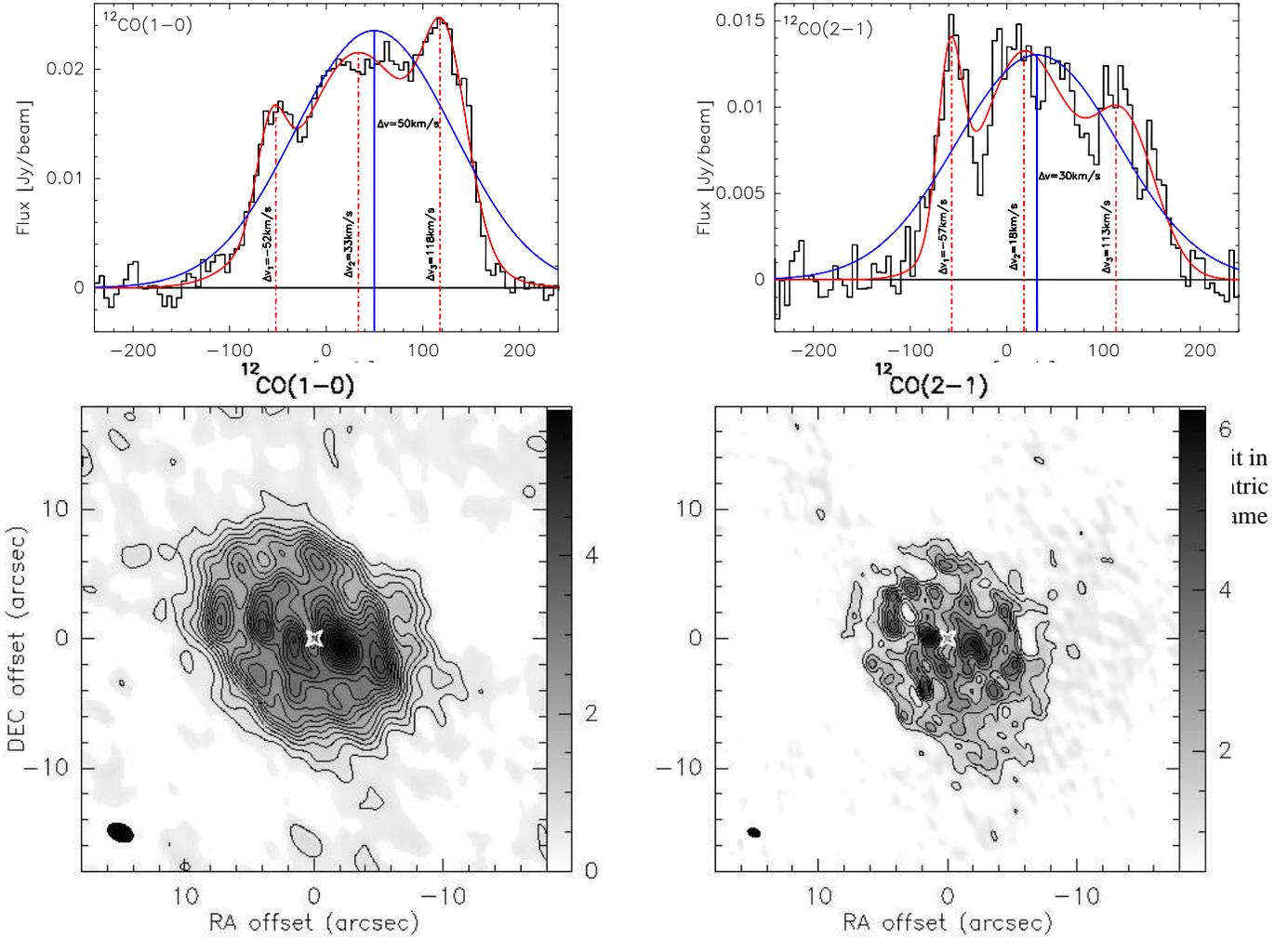}
\caption{\textit{Left panel}: $^{12}$CO(1--0) integrated intensity contours 
observed with the IRAM PdBI+30\,m toward the center of \nnn. 
The white star marks the coordinates 
of the dynamical center of the galaxy, with offsets 
in arcseconds.
The map, derived with 2$\sigma$ clipping, has not been corrected for
primary beam attenuation.
The {\it rms} noise level is $\sigma = 0.09\,{\rm Jy\,beam^{-1}\,km\,s^{-1}}$ 
and contour levels run from 3$\sigma$ to 21$\sigma$ with 3$\sigma$ spacing.
In this map the $\pm 130\,{\rm km\,s^{-1}}$ velocity range is used. 
The beam of 2\farcs1 $\times$ 1\farcs4 is plotted in the lower left.
\textit{Right panel}: Same for $^{12}$CO(2--1).
The {\it rms} noise level is $\sigma = 0.2\,{\rm Jy\,beam^{-1}\,km\,s^{-1}}$ and 
contour levels run from 3$\sigma$ to 10$\sigma$ with 3$\sigma$ spacing.
The beam of 1\farcs1 $\times$ 0\farcs7
is plotted at lower left.}
\label{fig:co10-21}
\end{figure*}

The $^{12}$CO(1--0) and $^{12}$CO(2--1) integrated intensity 
distributions are  shown in Figure \ref{fig:co10-21}.
The %$^{12}$CO(1--0) and $^{12}$CO(2--1) 
CO emission is distributed over a disk of $\sim$16\arcsec 
($\sim$2.2\,kpc) diameter.
Our $^{12}$CO(1--0) observations show several peaks, 
distributed more or less randomly, with the strongest one 
offset from the nucleus $\sim$2\arcsec\  
toward the west/southwest, away from NGC\,5954.
The $^{12}$CO(1--0) distribution is different from the \hi\ 
morphology, where the strongest emission has been detected mostly on 
the side nearer to NGC\,5954 \citep[e.g.,][]{iono05},
although the \hi\ resolution is probably insufficient to resolve
distinct peaks.
In the $^{12}$CO(2--1) map the central emission is also clearly resolved
and more clumpy than in $^{12}$CO(1--0).
The strongest $^{12}$CO(2--1) peak is not that at $\sim$2-3\arcsec\ 
in the west/southwest direction from the nucleus, like for 
$^{12}$CO(1--0), but that at $\sim$1\farcs5 in the east direction from the nucleus.

The total H$_{2}$ mass derived from our interferometric (PdBI+30\,m) 
map within the 42\arcsec\ primary beam field of the PdBI 
($S\rm_{CO} = 254$\,Jy km\,s$^{-1}$, see Table \ref{table2}) 
is M$\rm_{H_{2}}$$\sim$1.7$\times 10^{9}~\rm{M_{\odot}}$ 
(M$\rm_{mol}$$\sim$2.3$\times 10^{9}~\rm{M_{\odot}}$).
Within a diameter of 22\arcsec\ ($S\rm_{CO} = 185$\,Jy km\,s$^{-1}$, see 
Table \ref{table2}), we derive a H$_{2}$ mass of  
M$\rm_{H_{2}}$$\sim$1.3$\times 10^{9}~\rm{M_{\odot}}$, 
$\sim$50\% of the single dish 30\,m  H$_{2}$ mass (see Sect. \ref{sec:30m})
corresponding to a region about 6.6 times larger in area.
Hence, half of the molecular gas is concentrated in the central region
of \nnn, as expected for spiral galaxies.
The molecular gas distribution consists mainly of individual giant
molecular cloud complexes, with the biggest one alone having a mass of 
M$\rm_{H_{2}}$ = 3.0 $\times$ 10$^8~\rm{M_{\odot}}$, while the smaller ones 
have masses of a few 10$^{7}~\rm{M_{\odot}}$. 

\nnn\ is quite massive in molecular gas compared to
the other NUGA galaxies, where typically the mass is on the order
of $\sim$3$\times10^8~\rm{M_{\odot}}$. Three galaxies of the NUGA
sample, NGC\,4569 \citep{fred07}, NGC\,2782 \citep{leslie08},
NGC\,3147 \citep{vivi08}, with molecular masses in the range of
1-3$\times10^9~\rm{M_{\odot}}$, are comparable with \nnn.
The extraordinary case of NGC\,1961
\citep{francoise09} is excluded from these considerations: NGC\,1961 exhibits
a H$_{2}$ mass of $\sim$1.8$\times10^{10}~\rm{M_{\odot}}$, 
almost an order of magnitude more massive than 
any NUGA galaxy.

\subsection{CO line ratio \label{sec:coratios}}

\begin{figure}
\centering
\includegraphics[width=0.4\textwidth,angle=-90]{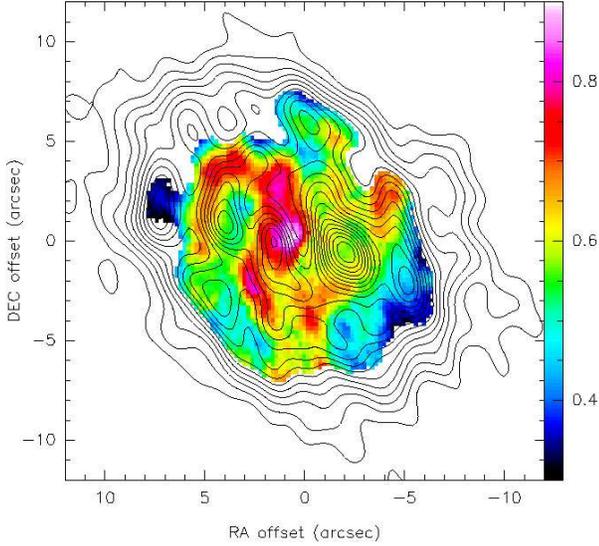}
\caption{
Color scale of the CO(2--1)/CO(1--0) ratio map 
and $^{12}$CO(1--0) intensity map contours as in Fig. \ref{fig:co10-21} 
(left panel).
}
\label{fig:ratio}
\end{figure}

The comparison between the two $^{12}$CO maps, obtained after
convolving the $^{12}$CO(2--1) map to the lower resolution of
the $^{12}$CO(1--0) one and including short spacings for  both
maps, gives information about the local excitation conditions
of the molecular gas.
%  First, we have checked if the two $^{12}$CO distributions
%  reported at the same resolution are in agreement. 
% When the $^{12}$CO(2--1) 
% data are tapered and convolved to the $^{12}$CO(1--0) resolution,
% the maxima of the $^{12}$CO(2--1) distribution
% agree quite well with those of $^{12}$CO(1--0).
% The partial agreement is visible for the peak at 
% $\sim$2$^{\prime\prime}$ toward southwest with respect to
Figure \ref{fig:ratio} shows the ratio of the  $^{12}$CO(1--0)
to $^{12}$CO(2--1) convolved to the same resolution with 
$^{12}$CO(1--0) contours as in Fig. \ref{fig:co10-21} (left panel).
The line ratio ranges from 0.3 to 0.9 inside the observed region 
and the bulk of the emission has a ratio between 0.4 and 0.8, 
values consistent with optically thick emission in spiral 
disks \citep[e.g.,][]{braine92,santi93}. 
Ratio values of $\sim$0.9 are reached in some clumpy structures, randomly 
distributed inside the mapped region, especially those toward E/NE with respect 
to the nucleus, well visible in $^{12}$CO(2--1) emission 
(Fig. \ref{fig:co10-21}, right panel).
A $R_{21}$ line ratio of 0.9 suggests a moderately high $^{12}$CO excitation
consistently with \citet{yao03} who have detected the $^{12}$CO(3--2) 
line (345.80 GHz) with an intensity of $I_{32}=38.7$ K km s$^{-1}$ 
and a line ratio $R_{32}=I_{32}/I_{10}$=0.85 in the nucleus of 
\nnn.
%Toward the south, the line ratio apparently reaches values of $\sim$1.2-1.3; 
%this could be an effect of convolution.
Different excitation conditions of the molecular gas appear to 
characterize the interacting companion galaxy NGC\,5954,
where the $^{12}$CO(3--2) line has been detected with 
%an intensity of $I_{32}$=14.4 K km s$^{-1}$ and 
% is a smaller galaxy, so we would expect less flux
a ratio of $R_{32}$=0.37 \citep{yao03}.

\subsection{Kinematics \label{sec:cokinematics}}
The velocity-channel maps (see Figs. \ref{channels10} 
and \ref{channels21}) 
show a general regularity of the large scale kinematics, 
typical for a rotating disk.
Some local wiggles, not forming a coherent grand design, are 
superimposed on this regular pattern, both at negative and positive 
velocities especially to the west of the nucleus.
These kinematic ``glitches'' are probably associated with the
intensity peak toward the SW.

% The velocity-channel maps 
% %of the $^{12}$CO(1--0) and $^{12}$CO(2--1) emissions respectively
% % in the central region of NGC\,5953.
% (see Figures \ref{channels10} and \ref{channels21}) 
% show a general regularity of the large scale kinematics, 
%  typical for a rotating disk. However, there are also some local wiggles 
%  (i.e., streaming motions) superimposed on this regular pattern 
%  both at negative and positive velocities, almost exclusively to the 
%  west of the nucleus.
% These kinematic ``glitches'' are probably associated with the
% intensity peak toward the SW.

\begin{figure}
\centering
\includegraphics[width=0.4\textwidth,angle=-90]{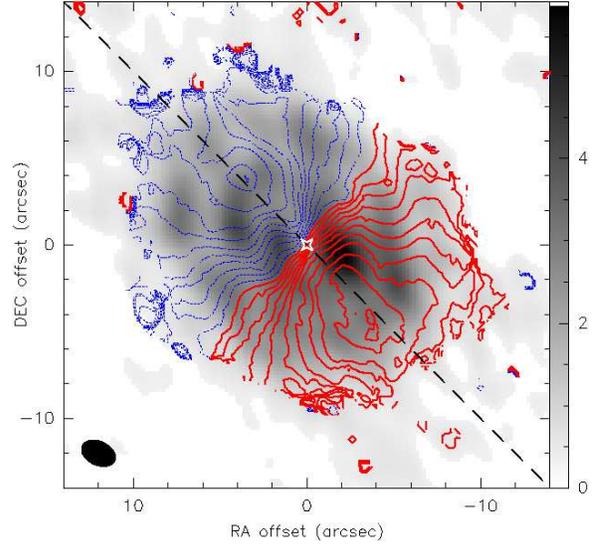}
\caption{Overlay of the integrated $^{12}$CO(1--0) emission, same as 
Fig. \ref{fig:co10-21} (left panel), with CO mean-velocity field 
in contours spanning the range -100 to 100 km s$^{-1}$ 
in steps of 10 km s$^{-1}$. 
The white star indicates the dynamical center of the galaxy.
The velocities are referred to $V_{\rm sys, hel}$ = 1990 km\,s$^{-1}$, 
solid (red) lines are used for positive velocities, and 
dashed (blue) lines for negative velocities.
The dotted line indicates the position angle of the major axis 
(PA = 45$^{\circ}$ $\pm$ 1$^{\circ}$).}
\label{fig:co10-velo}
\end{figure}

Figure \ref{fig:co10-velo} shows the 
$^{12}$CO(1--0) isovelocity contours (first-moment map) superposed 
on the $^{12}$CO(1--0) integrated intensity.
The white star indicates the dynamical center 
of the galaxy assumed coincident with the 
phase tracking center of our observations, and the velocities
are relative to the systemic heliocentric velocity, 
$V_{\rm sys, hel}$ = 1990 km\,s$^{-1}$.
The dotted line traces the major axis of the galaxy determined
from our observations (PA = 45$^{\circ}$ $\pm$ 1$^{\circ}$),
by maximizing the symmetry in the position velocity diagrams.
This value is also the best-fit PA given by the elliptically-averaged
surface brightness profiles discussed in Sect. \ref{sec:otherdata}.
%  The velocity field resembles one of a non-perturbed spiral disk, as
%  found by \citet{hernandez03} for the {\nii} velocity field.

\begin{figure}[h]
\centering
\includegraphics[width=0.6\textwidth,angle=-90]{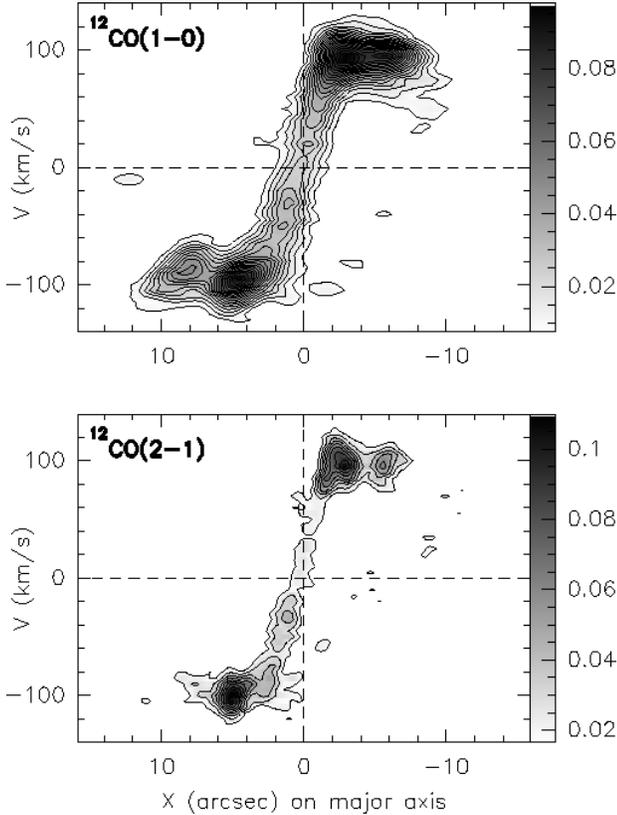}
\caption{\textit{Upper panel}: $^{12}$CO(1--0) position-velocity
diagram along the major axis (PA = 45$^\circ$) of \nnn\
using the velocity range from -140 to 140 km s$^{-1}$ contoured 
over a grey-scale representation.
Contour levels are from 3$\sigma$ to 39$\sigma$ in steps of
2$\sigma$ ($\sigma$=2.5\,mJy\,beam$^{-1}$). 
The velocities are relative to $V_{\rm sys, hel}$ (=1990 km\,s$^{-1}$) and 
X are the offsets along the major axis in arcsecs.
\textit{Bottom panel}: The same for $^{12}$CO(2--1). 
Contour levels are from 3$\sigma$ to 20$\sigma$ 
in steps of 2$\sigma$ ($\sigma$=5.5\,mJy\,beam$^{-1}$).
}
 \label{p-v-major}
 \end{figure}

\begin{figure}[h]
\centering
\includegraphics[width=0.6\textwidth,angle=-90]{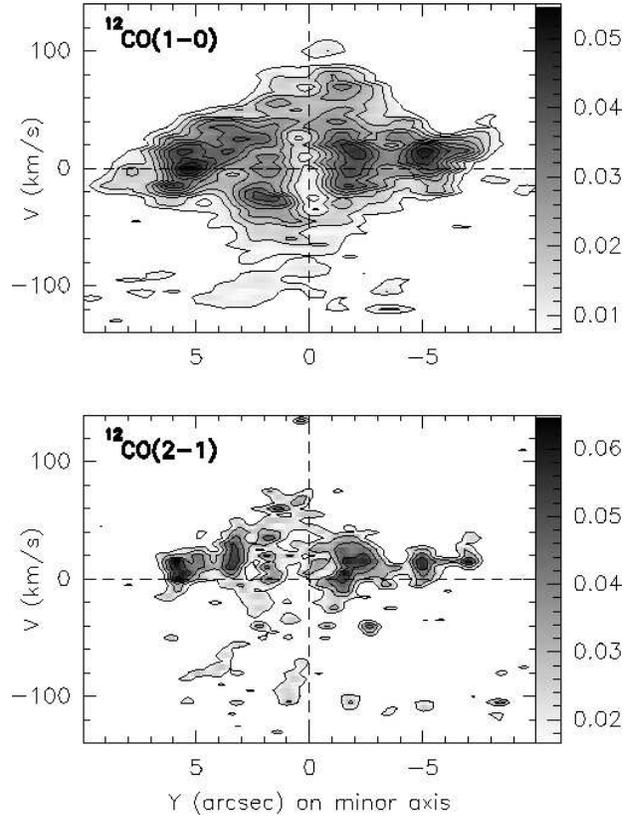}
\caption{\textit{Upper panel}: 
Same as Fig.\ref{p-v-major} along the minor axis 
(PA = 135$^\circ$) of \nnn.
Contour levels are from 3$\sigma$ to 22$\sigma$ in steps of
2$\sigma$.
Y are the offsets along the minor axis in arcsecs.
\textit{Bottom panel}: The same for $^{12}$CO(2--1). 
Contour levels are from 3$\sigma$ to 12$\sigma$ in steps of
2$\sigma$.
}
\label{p-v-minor}
\end{figure}

Position-velocity (p-v) cuts along the major (PA = 45$^\circ$) and 
minor axis (PA = 135$^\circ$) of \nnn\ are shown in 
Figures \ref{p-v-major} and \ref{p-v-minor}, respectively.
In both figures, the $^{12}$CO(1--0) emission is given in the top panel 
and the $^{12}$CO(2--1) one in the bottom panel.
The p-v diagrams along the major axis 
reveal regular circular rotation and appear symmetric 
in the inner 10-11$^{\prime\prime}$, 
both in $^{12}$CO(1--0) and $^{12}$CO(2--1).
This regularity in the p-v diagram of \nnn, more typical for 
normal spiral galaxies than interacting ones, has also been found 
in the ionized gas, including H$\alpha$, 
{\oiii} \citep{gonzalez96}, and {\nii} \citep{hernandez03}.
The kinematics shown by the minor-axis p-v diagrams
(Fig. \ref{p-v-minor}) are also quite regular, but show significant 
velocity dispersion for the $^{12}$CO(1--0) close to the
nucleus. This effect can be  attributed to beam smearing.

\subsection{The rotation curve and dynamical mass\label{sec:dynmass}}

We have derived a rotation curve (RC) from the p-v diagram taken along
the kinematic major axis of \nnn\ at a PA = 45\,$^\circ$. 
The terminal velocities were derived by fitting multiple gaussian
profiles to the spectra across the major axis. 
The fitted velocity centroids, corrected for $\sin i$ ($i\,=\,42^\circ$),
give $V_{\rm obs}/\sin i$ for each galactocentric distance. 

\begin{figure*}
\centering
\hbox{
\includegraphics[width=0.50\textwidth,bb=34 296 588 648]{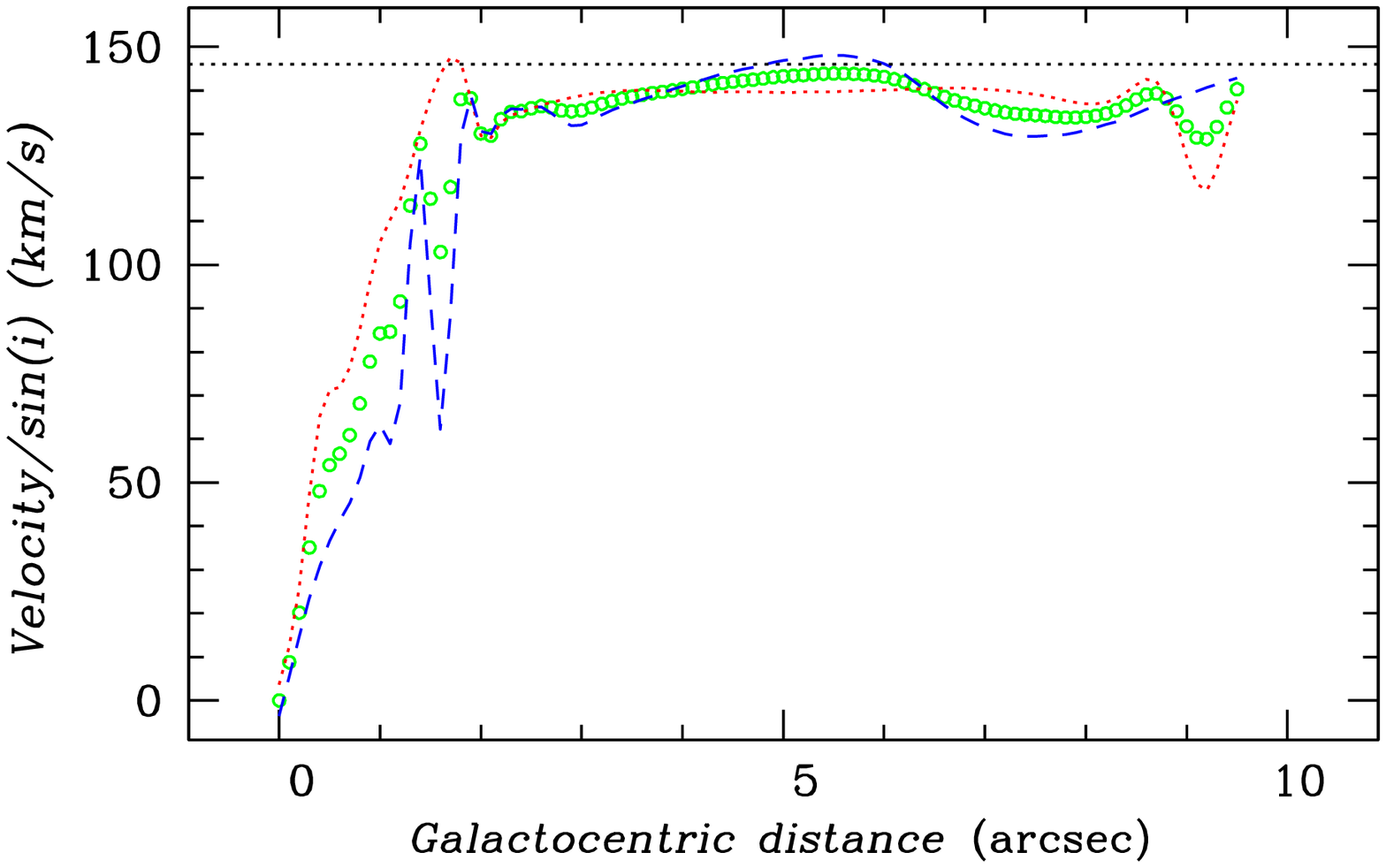}
\hspace{-0.099\textwidth}
\includegraphics[width=0.50\textwidth,bb=34 296 588 648]{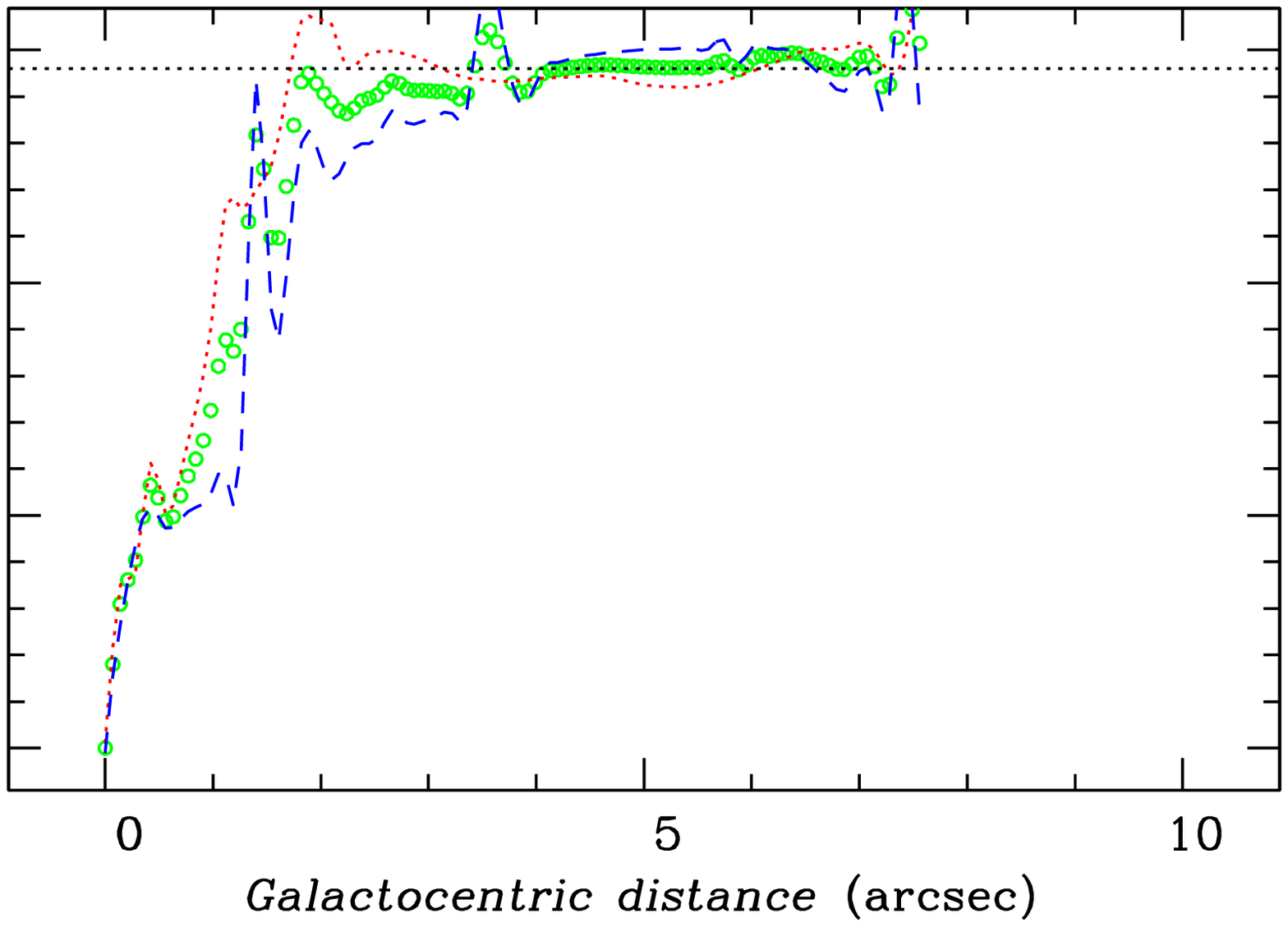}
}
\caption{\textit{Left panel}:
The $^{12}$CO(1--0) rotation curve derived as described in the text.
The positive (negative) velocities are shown as a red dotted (blue dashed)
line; the open (green) circles show the average. The horizontal dotted 
line at 146 km s$^{-1}$ indicates the velocity of the bulk of the molecular 
gas.
\textit{Right panel}:
The same as the left panel, but for $^{12}$CO(2--1).
}
\label{fig:rc}
\end{figure*}

For both lines the velocities (negative) to the northeast 
increase less steeply within $\sim$250\,pc than those to the south (positive
velocities).
Nevertheless, they converge to the same value at greater distances, so
we averaged together the two curves derived from either side of the major axis. 
%The resulting \vrot\ may be slightly shallower in the
%inner regions than the true mass distribution would imply.
Because data for both lines were consistent, we combined both curves
into an average by spline interpolation.
The RCs from the two $^{12}$CO transitions are shown in Figure \ref{fig:rc}. 
They are consistent with those found for the ionized gas,
as reported by \citet{gonzalez96} and \citet{hernandez03}.

The $^{12}$CO RCs are very regular and apparently this behavior 
is expected more for normal spiral galaxies than interacting systems.
\nnn\ and the companion are clearly interacting and their outer 
parts are perturbed as shown by the presence of a tidal tail seen both in optical 
and in \hi\ emission \citep[e.g.,][]{hernandez03,haan08}.
In \nnn, we are mapping the molecular gas in the inner parts 
that have a dynamical time-scale much shorter than the outer ones, 
and therefore they had the time to relax and reach an equilibrium 
state, perhaps after gas accretion/exchanges from/with NGC\,5954.  
In other words, the regularity of the RCs in $^{12}$CO is not unexpected.

%The apparent turnover at the end of the rotation curves may be
% real, although upon a more careful comparison the optical curves
% turn down in the negative velocities toward the NE, while
% we find a turndown in the positive velocities toward the SW.

In the major-axis p-v diagrams, the velocity peaks at 
$\sim$$120/(\sin i$)\,km\,s$^{-1}$ at a radius of $\sim$2--4\arcsec\
($\sim$400\,pc); however, the bulk of the gas is rotating
at $\sim$$98/(\sin i$)\,km\,s$^{-1}$ (see Fig. \ref{fig:rc}).
This is roughly consistent with what is found for the ionized gas,
$\sim$$115/(\sin i$)\,km\,s$^{-1}$ by \citet{hernandez03}\footnote{Taking into
account their different adopted inclination.}.
The bulk velocity corrected for disk inclination, $\sin\,i$ (we adopt 
$i=42^\circ$, see Sect. \ref{sec:dyncen}), becomes $\sim$146\,km\,s$^{-1}$. 
From the corrected bulk velocity we can estimate the dynamical mass 
within a certain radius using the formula 
$M(R) = \alpha \times R V^2(R)/G$,
where $M(R)$ is in $\rm{M_{\odot}}$, $R$ in kpc, and $V$ in km\,s$^{-1}$.
Assuming the most flattened disk-like distribution ($\alpha = 0.6$),
the above formula gives a dynamical mass of 
M$_{\rm dyn}=2.4\times10^{9}$\,$M_\odot$
within a radius of 6\arcsec\ ($\sim$0.8\,kpc).

Assuming a roughly flat rotation curve, the dynamical mass 
should be M$_{\rm dyn}=8.6\times10^{9}$\,$\rm{M_{\odot}}$
within a radius of 21\arcsec\ ($\sim$2.9\,kpc). 
In the same region, we estimate a total molecular gas mass 
of $\sim$1.7$\times10^{9}$\,$\rm{M_{\odot}}$ (see Sect. \ref{sec:comorphology}),
a fraction of $\sim$$20$\% of the dynamical mass.
%\citet{iono05} estimated an even higher fraction, $\sim$35\%.
%This extraordinarily high gas mass fraction will be discussed 
%in Sect. \ref{sec:discussion}.
%One of the main results is the high molecular gas mass fraction ($\sim$22$\%$) 
%with respect to the dynamical mass found in the inner 
%22\arcsec\ ($\sim$3\,kpc). 
Anyway, this percentage is subject to uncertainties.
One of these could be due to the variation of the 
H$_{2}$--CO conversion factor.
In this work, as for similar studies of molecular gas, we assume
a constant value for the $X$ conversion factor, for instance that determined 
for the Milky Way.
However, the $X$ ratio is determined by various factors, such as the metallicity, 
the temperature, the cosmic ray density, and the UV radiation field 
\citep[see][]{maloney88,boselli02}, and its value can change by a 
factor 4-15 \citep[e.g.,][]{wilson95,vivi07}. 
The $X$ value also changes with galaxy morphological type:
usually galaxies earlier than Scd show values comparable to, or lower
than, the Galactic one, while extremely late-type spirals or irregular
galaxies tend to show higher values \citep[][]{nakai95}.
Another possible uncertainty, but on the estimate of the dynamical mass,
could be the suspected warp of the disk, discussed later 
in Sect. \ref{sec:ring}, that might give a wrong inclination of the disk, 
and so an incorrect rotation velocity.
%This percentage of gas mass could result from the interaction 
%and ensuing gas accretion.
%Another possible explanation could be the variation of the 
%H$_{2}$--CO conversion factor.
%In this work, as for similar studies of molecular gas, one assumes
%a constant value for the $X$ conversion factor, for instance that determined 
%for the Milky Way.
%However, the $X$ ratio is determined by various factors, such as the metallicity, 
%the temperature, the cosmic ray density, and the UV radiation field 
%\citep[see][]{maloney88,boselli02}, and its value can change by a 
%factor 4-15 \citep[e.g.,][]{wilson95,vivi07}. 
%The $X$ value also changes with galaxy morphological type:
%usually galaxies earlier than Scd show values comparable to, or lower
%than, the Galactic one, while extremely late-type spirals or irregular
%galaxies tend to show higher values \citep[][]{nakai95}.

\section{Dust, stellar structure, and star formation\label{sec:stars}}
% \section{Comparison with other wavelengths\label{sec:comparison}}

In this section we compare the $^{12}$CO distribution of 
\nnn\ with observations at other wavelengths.
These comparisons have a dual aim.
First, they allow the study of possible correlations between different
tracers of the ISM, such as between molecular clouds,
considered the birth site of future generations of stars,
and the light from the stellar component.
The second aim is to determine the location of dynamical resonances, 
which greatly aid the determination of gas flow in the
circumnuclear regions of the disk. However, in the case of \nnn, we 
find no evidence for any non-asymmetric component which could drive 
resonances.

\subsection{Optical and near-infrared morphology \label{sec:othermorphology}}

\begin{figure*}
\centering
\includegraphics[width=0.43\textwidth,angle=-90]{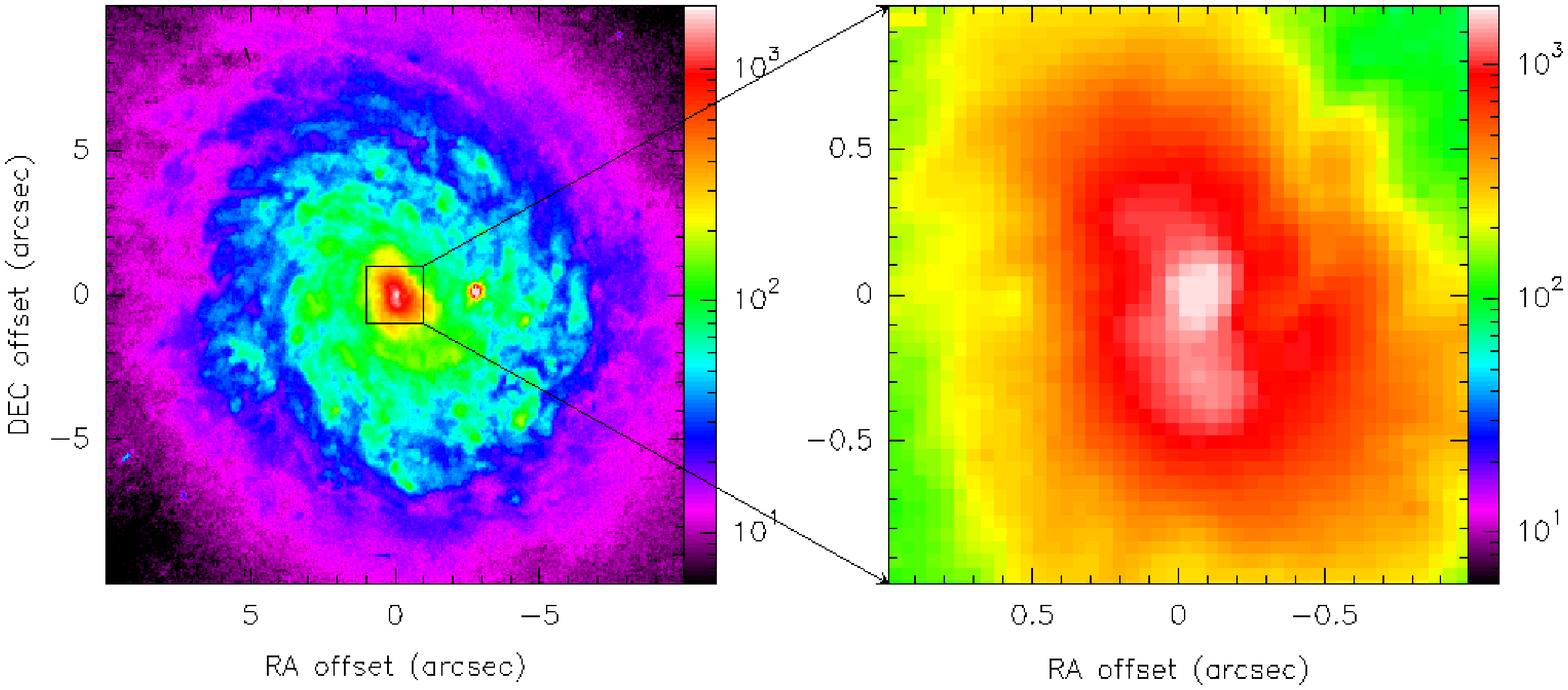}
\caption{\textit{Left panel}:
F606W band WFPC2/\hst\ image of  \nnn.
The inner $20^{\prime\prime}$ are shown.
\textit{Right panel}:
A zoom of the nuclear region.
The inner $2^{\prime\prime}$ are shown.
}
\label{fig:f606w}
\end{figure*}

Figure \ref{fig:f606w} (left panel) displays a \hst/NICMOS/F606W 
band image of the inner $20^{\prime\prime}$ of \nnn, 
described in Sect. \ref{sec:otherdata}.
We imposed that the brightness peak in this image coincides
with the phase tracking center of our $^{12}$CO observations
and therefore with the dynamical center, as discussed
in Sect. \ref{sec:dyncen}.
The small peak at $\sim$3\arcsec\ west of center is almost certainly the 
foreground star identified in \nnn\ by \citet{piraf90}.
Inspection of this figure shows a clearly-defined non-axisymmetric 
structure of $\sim$250\,pc in size,
similar in shape and size to the [O{\sc iii}] emission found by \citet{gonzalez96}. 
A close-up of the inner $2^{\prime\prime}$ (Fig. \ref{fig:f606w}, right panel)
reveals an ``S-shape'' feature, perhaps associated with a nuclear
bar \citep[e.g.,][]{hernandez03}.
This bar-like signature is $\simgt$60\,pc in radius,
with a PA$\sim$17$^\circ$.
%almost vertical in the center at PA$\sim$0$^{\circ}$.
\citet{hernandez03}, by examining a contrast-enhanced $B$-band 
image of \nnn\ (the same image as used here),
identified this structure
at about the same PA as the global major axis of the galaxy.
We find (see Fig. \ref{fig:f606w}, right panel) 
instead a more vertical orientation, PA$\sim$$5-20^\circ$,
depending on which features of the S-shape are used to define the PA.
Rather than a nuclear bar, the S-shape structure
could also be an optical counterpart of the radio jet \citep{melanie07a},
because of the similar morphology, position angle, and size;
this feature also seems to correspond to a higher excitation of the ionized
gas \citep{gonzalez96}.
  
\begin{figure*}
\centering
\includegraphics[width=0.43\textwidth,angle=-90]{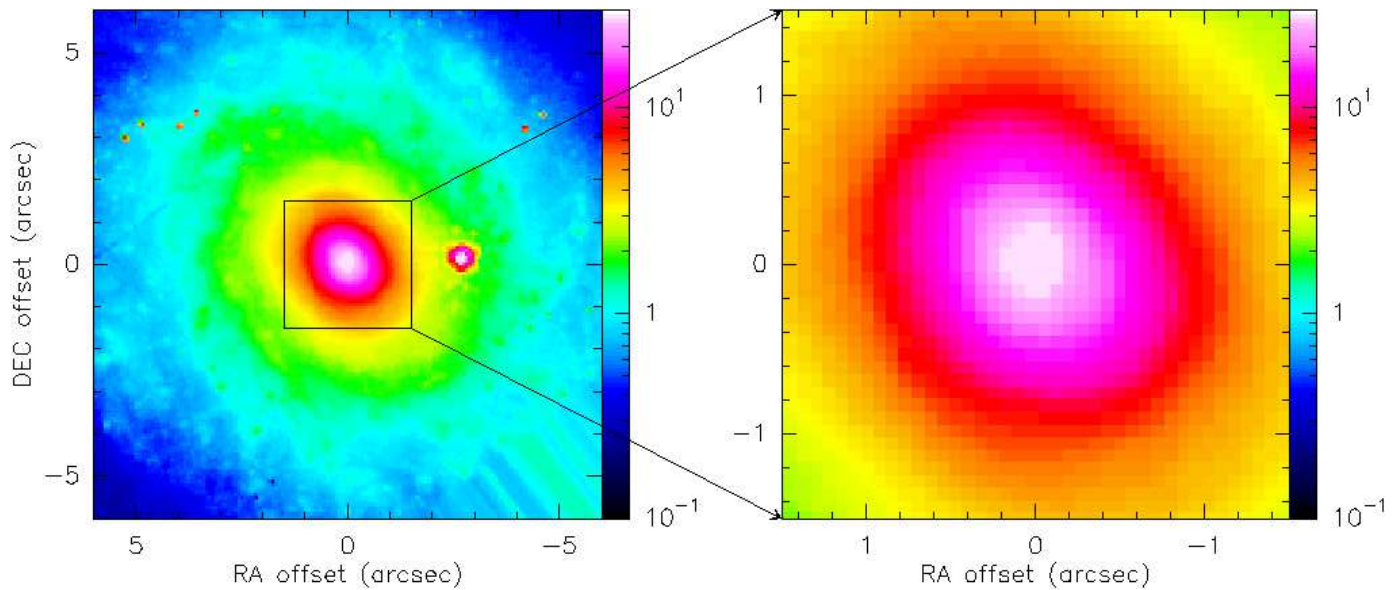}
\caption{\textit{Left panel}: F160W band NICMOS/\hst\ 
image of \nnn.
The inner $12^{\prime\prime}$ are shown.
\textit{Right panel}:
A zoom of the nuclear region.
The inner $3^{\prime\prime}$ are shown.
}
\label{fig:f160w}
\end{figure*}

The inner $1.6\,{\rm \mu m}$ morphology 
(\hst/NICMOS/F160W) of \nnn\ is shown in Figure \ref{fig:f160w}.
The foreground star to the 3\arcsec\ to the west of the nucleus
is clearly visible also at $1.6\,{\rm \mu m}$.
The flocculent spiral structure clearly visible in the optical 
(Fig. \ref{fig:f606w}) is not seen in the NIR emission; 
there is also no NIR counterpart to the S-shape feature seen
in the optical.
%There is no evidence either for the oval feature or for the S-shape
%structure in the elliptically averaged surface brightness profiles
%(see Fig. \ref{fig:profiles}).
%This could be because of the coincidence of the position angles of
%the morphological features and the line of nodes,
%which renders more difficult the identification of non-axisymmetric structure.

\begin{figure*}
\centering
\includegraphics[width=0.43\textwidth,angle=-90]{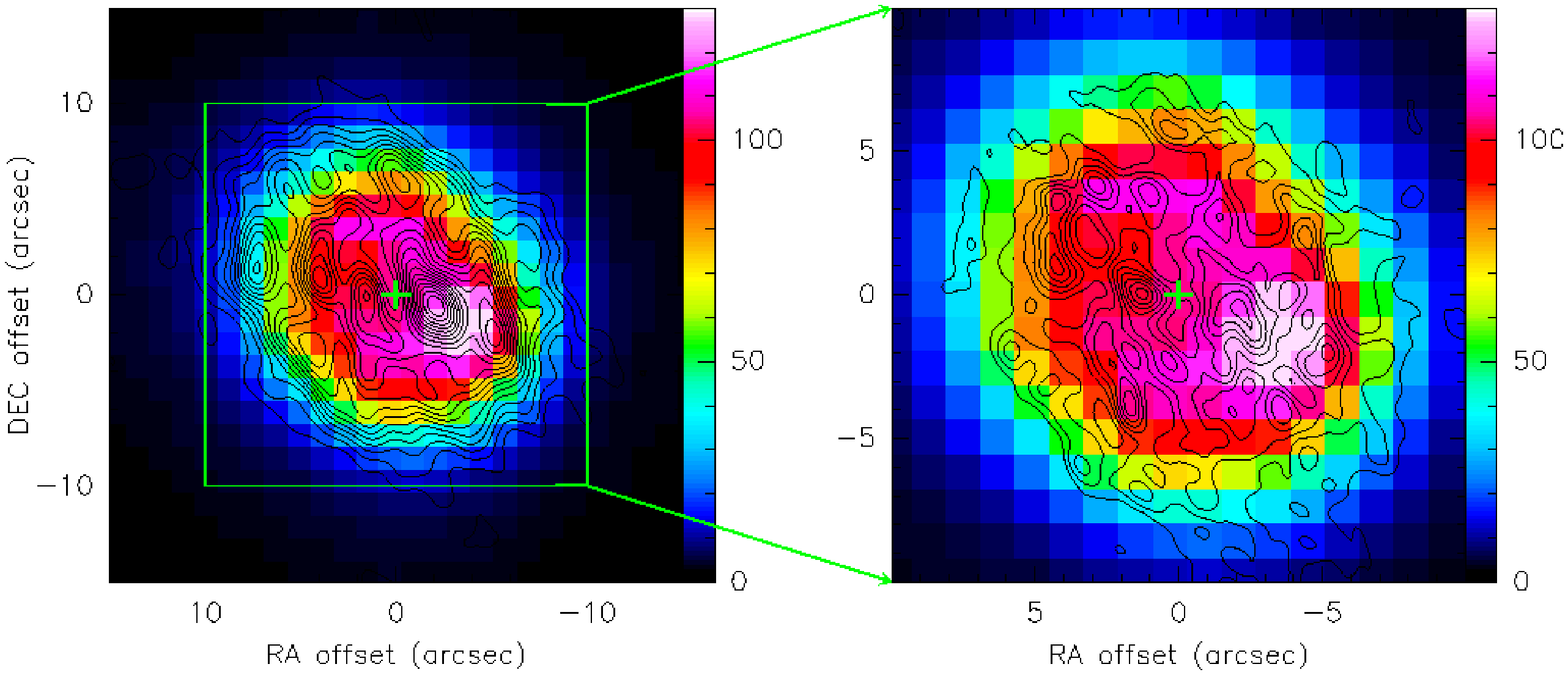}
\caption{\textit{Left panel}: 
$^{12}$CO(1--0) integrated intensity contours as in 
Fig. \ref{fig:co10-21} (left panel) overlaid on the IRAC 8.0\,$\mu$m 
``dust-only'' image of \nnn\ in false color. The (green) cross marks the 
dynamical center. The central 30$^{\prime\prime}$ are shown.
\textit{Right panel}: Same for $^{12}$CO(2--1) integrated intensity contours 
as in Fig. \ref{fig:co10-21} (right panel). The central 20$^{\prime\prime}$ are 
shown.
% IRAC 8.0\,$\mu$m ``dust-only'' image of NGC\,5953 in false color,
% with CO(1--0) [CO(2--1)] contours overlaid  in the left (right) panel.
% The cross marks the dynamical center.
}
\label{fig:dustonly}
\end{figure*}

The morphology of the warm dust in the circumnuclear regions
of \nnn\ is shown in Figure \ref{fig:dustonly}, where $^{12}$CO(1--0) and
$^{12}$CO(2--1) intensities are contoured over the dust-only image described in
Sect. \ref{sec:otherdata}.
There is a dust emission peak to the SW, not exactly coincident with the 
$^{12}$CO emission peak but roughly in the same direction.
The bulk of the (putative) dust emission is configured in a smooth featureless disk,
similar to that seen in the $^{12}$CO emission.

\subsection{Stellar structure\label{sec:stellar}}

\begin{figure}
\centering
\includegraphics[width=0.45\textwidth,angle=-90]{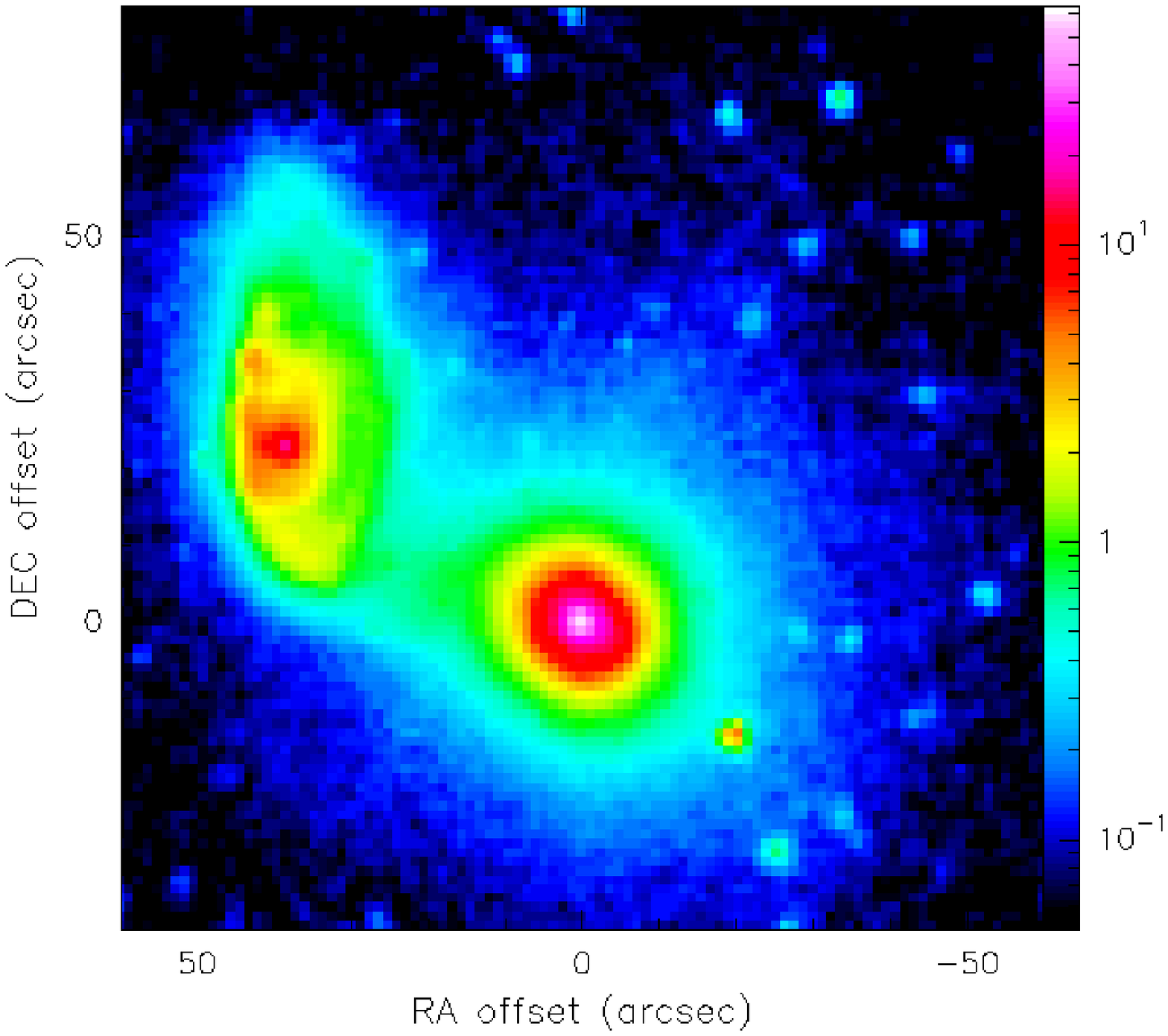}
\caption{
IRAC 3.6\,$\mu$m image of \nnn\ and NGC\,5954, coordinate system centered on \nnn\ 
(with N up, E left).
The inner $120^{\prime\prime}$ are shown.
}
\label{fig:3.6mu}
\end{figure}

A large-scale view of the interacting companions \nnn\ and  
NGC\,5954 at 3.6\,$\mu$m is illustrated in Figure \ref{fig:3.6mu}.
Like the $K$ band,
this image traces the massive component of the stellar populations in galaxies,
but at 3.6\,\micron\ the extinction is about 1/3 that of the $K$ band
\citep{cardelli89}.

To investigate stellar structure, 
we performed a two-dimensional B/D decomposition with
\textit{galfit} \citep{peng02} on
both the ground-based $H$-band image and IRAC 3.6\,$\mu$m image, 
fitting both galaxies (NGC\,5953 and NGC\,5954) simultaneously.
In the fit, the models were convolved with the IRAC Point Response Function (PRF)
or an approximation of the ground-based Point Spread Function, as appropriate.
The best fit included a Sersic generalized exponential bulge, an 
exponential disk, and a nuclear point source for \nnn, and a 
Sersic generalized exponential bulge (but with index $n\sim$1, 
so this is probably in reality a disk) and a nuclear point source 
for NGC\,5954. 
The IRAC fit has a reduced $\chi^2_\nu$ of 4.2 (averaged over \nnn\
and NGC\,5954), roughly ten times smaller than that in the $H$-band.
Hence, in what follows, we mainly rely on the IRAC fit.
The bulge comprises more than half ($\sim$59\%) of the total 3.6\,\micron\
light, consistently with its early Hubble type,
but has a small ($n\sim1$) Sersic index, unlike most early-type
bulges with $n\sim3$ \citep{moriondo98}.
The disk scalelength is $\sim$4\,kpc, slightly larger than typical
early-type spiral disks \citep[e.g.,][]{leslie04a};
this could be due to the interaction and the proximity with NGC\,5954.

\begin{figure*}
\centering
\includegraphics[width=0.43\textwidth,angle=-90,bb=145 17 460 750]{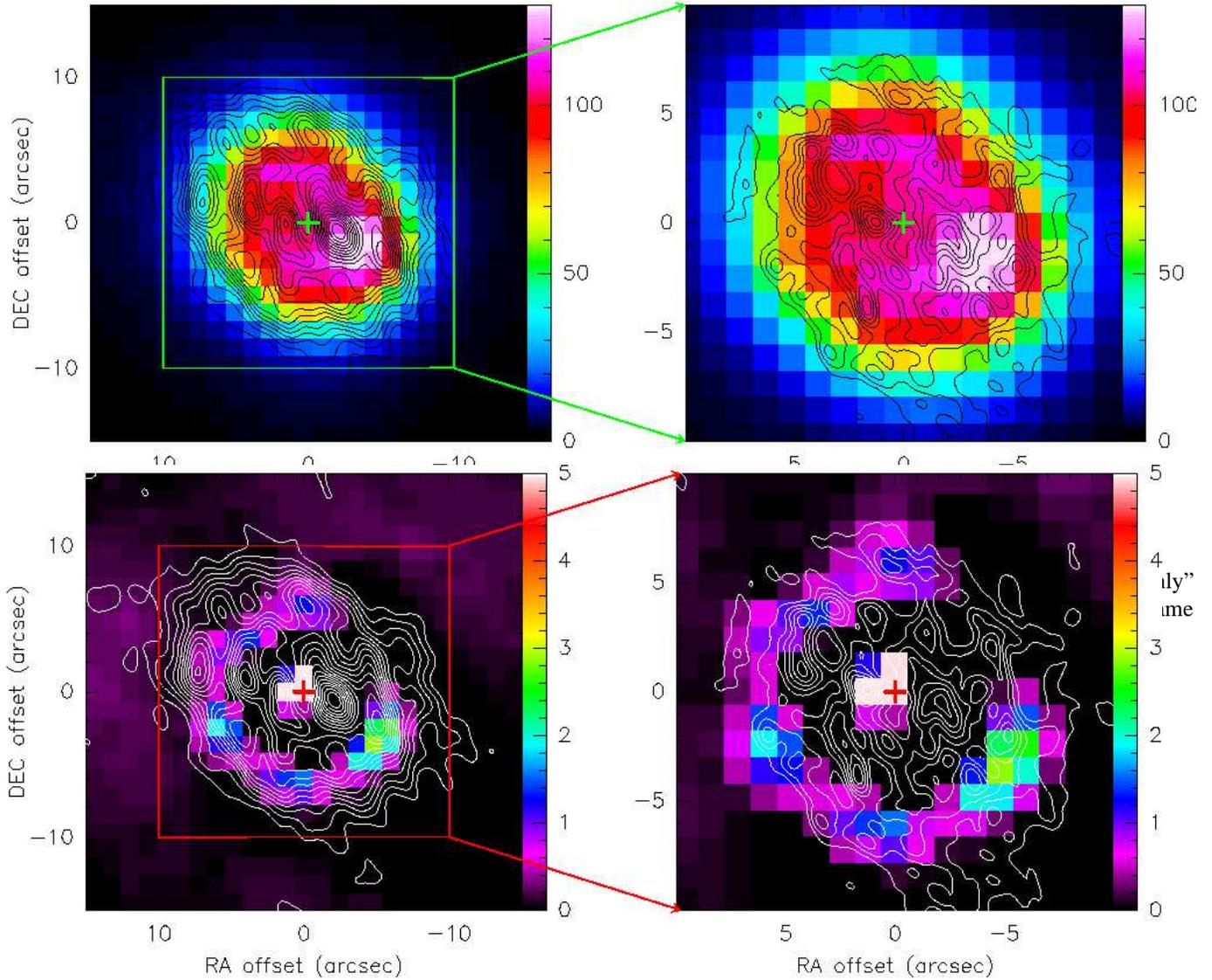}
\caption{\textit{Left panel}: 
$^{12}$CO(1--0) integrated intensity contours as in Fig. \ref{fig:co10-21} 
(left panel) overlaid on the residuals 
(in MJy\,sr$^{-1}$) of the B/D decomposition performed on the IRAC 3.6\,$\mu$m image.
The (red) cross marks the dynamical center.
The central 30$^{\prime\prime}$ are shown.
\textit{Right panel}: Same for $^{12}$CO(2--1) integrated intensity contours 
as in Fig. \ref{fig:co10-21} (right panel).
The central 20$^{\prime\prime}$ are shown.
}
\label{fig:res-co}
\end{figure*}

The IRAC 3.6\,\micron\ residuals from the B/D decomposition
are shown in Figure \ref{fig:res-co}, with the $^{12}$CO(1--0) 
(left panel) and the $^{12}$CO(2--1) (right panel) in contours.
The residuals show a low-amplitude circumnuclear ``ring'' with a 
radius of $\sim$5-7\arcsec.
The $^{12}$CO(1--0) and $^{12}$CO(2--1) disks nearly coincide 
in size with the ring structure in the residuals.

\begin{figure*}
\centerline{
\hbox{
\includegraphics[width=0.4\textwidth,angle=-90]{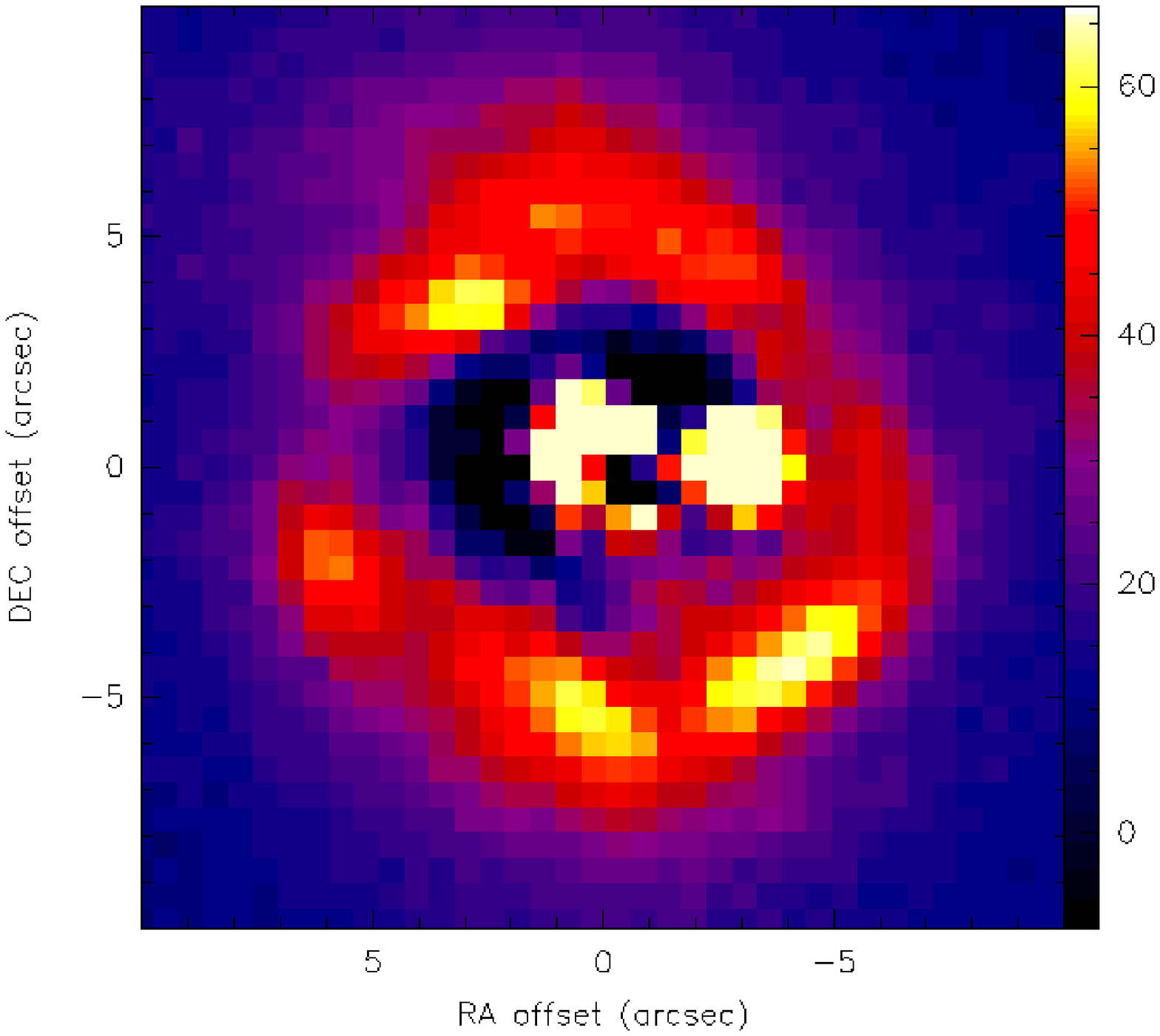}
\hspace{0.1\textwidth}
\includegraphics[width=0.4\textwidth,angle=-90]{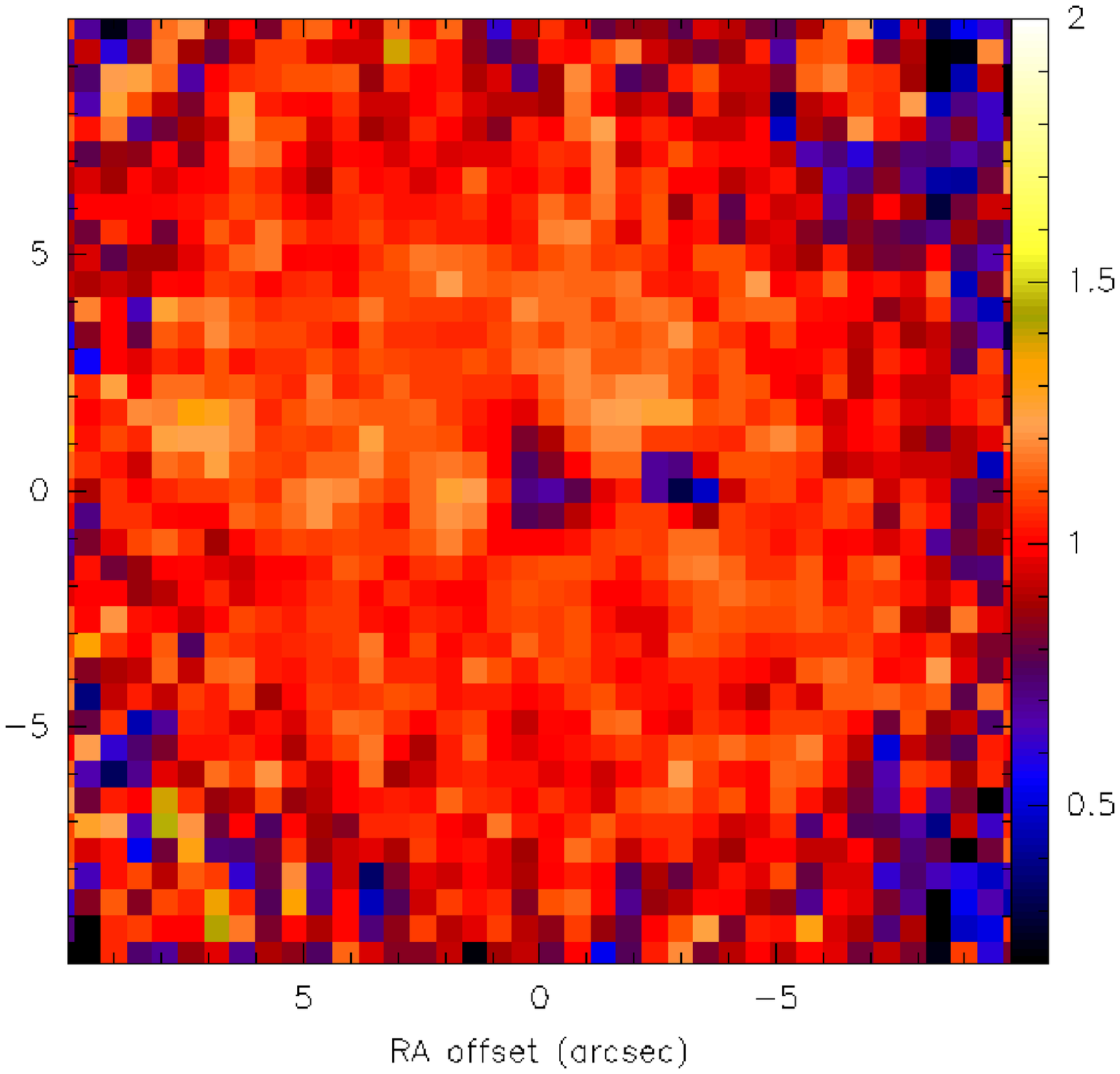}
}
}
\caption{
\textit{Left panel}: $H$-band residuals from the \textit{galfit} B/D decomposition.
\textit{Right panel}: The $J-K$ color image.
The FOV for both panels is 20\arcsec\ with north up, and east to the left.
In both images, the foreground star $\sim$2\farcs7 to the west is evident.
}
\label{fig:hres-jk}
\end{figure*}

Because the size of the ring in the IRAC residuals is typical
of small-scale ``doughnuts'' associated with the diffraction limit and
PRF incompatibilities, we wanted to verify the reality of
the ring-like feature seen in the 3.6\,\micron\ residuals.
If real, it should also appear in the $H$-band residuals, which
are shown in the left panel of Figure \ref{fig:hres-jk}.
Inspection of the figure shows that the ring is indeed present,
delineated by a series of surface brightness excesses.
The intensity contrast in these is around a factor of 10
relative to the regions external to the ring.
The presence of the ring in both sets of residuals suggests that it
is a real feature, rather than an artefact of IRAC PRF mismatching. 
Moreover, intensity peaks in the ring-like structure are also present 
in the un-sharp masked \hst/NICMOS/F160W map \citep[see][]{leslie04b}.
We therefore conclude that the ring in the NIR residuals is
real, and probably stellar in nature, as we discuss below.

\subsection{The stellar ring \label{sec:ring}}

The right panel of Fig. \ref{fig:hres-jk} displays the ARNICA/NOT $J-K$ 
color image of \nnn. %of both interacting galaxies, centered on NGC\,5953.
The nucleus of \nnn\ and a small region to the west have
extremely blue $J-K$ colors ($\sim$0.7 and $\sim$0.3, respectively), while
the disk has rather normal ones ($\sim$1.0).
The small knot 3\arcsec\ west of center, visible also
in the WFPC2/F606W (Fig. \ref{fig:f606w}) and NICMOS/F160W 
(Fig. \ref{fig:f160w}) images, coincides almost 
certainly with the star identified by \citet{piraf90}.
There may be a few isolated regions ($\sim$1\arcsec)
to the east of the nucleus with a red excess ($J-K$$\sim$1.3), but
there is no evidence in the $J-K$ colors for the ring seen in 
the B/D residuals.
This suggests that the excess residual ring is a stellar
ring, with colors similar to the surrounding disk.

\begin{figure}
\centering
\includegraphics[width=0.45\textwidth,angle=-90]{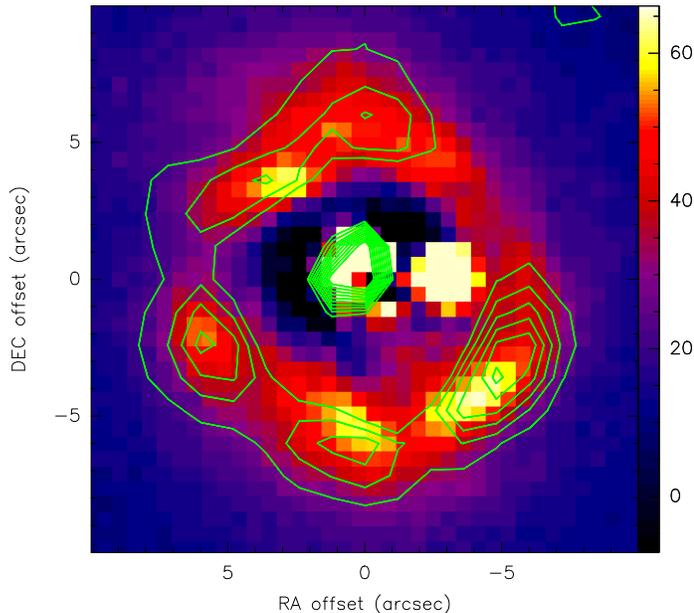}
\caption{
3.6\,$\mu$m residuals in (green) contours
(Fig. \ref{fig:res-co}) overlaid on the 
$H$-band residuals (Fig. \ref{fig:hres-jk}, left panel).
The inner $20^{\prime\prime}$ are shown.
}
\label{fig:tworings}
\end{figure}

A hint of the circumnuclear ring in \nnn\ also appears in 
previous optical observations of the ionized gas.
%H$\alpha$ observations \citep{hernandez03,falcon06}.
%The H$\alpha$ images of \nnn, obtained with
%the 1.5\,m OAN/SPM\footnote{Observatorio Astron\'{o}mico Nacional at San Pedro 
%M\'{a}rtir B. C. M\'{e}xico.}
%telescope \citep{hernandez03} and
With the integral-field spectrograph SAURON operating on the 
William Herschel Telescope, \citet{falcon06} found 
recent star formation in a circumnuclear ring with a radius of $\sim$6--7\arcsec.
The H$\beta$ emission is apparently distributed in a circumnuclear
disk, roughly coincident with the CO one,
but the H$\beta$ velocity dispersion and the [O{\sc iii}]/H$\beta$
ratio are {\it lower} in a ring-like structure than in the main circumnuclear disk.
This could be because the nucleus excites gas in an extended region \citep{falcon06},
or be due to a true ring in the dynamical sense of a resonance.
Lower velocity dispersion and lower [O{\sc iii}]/H$\beta$ are both
associated with enhanced star formation, leading \citet{hernandez03}
to conclude that \nnn\ hosts a circumnuclear ring.
The size of this star-forming ring is coincident with
that of the NIR ring in the residuals. 
Figure \ref{fig:tworings} shows the superposition of the 
3.6\,$\mu$m residuals image (Fig. \ref{fig:res-co}, left panel), 
represented in (green) contours, on the $H$-band 
residuals image (Fig. \ref{fig:hres-jk}, left panel). 
This figure clearly shows the coincidence in size 
between the two NIR rings, the same size of the 
star-forming ring seen in the ionized gas.

Because the ring is visible in ionized gas, but in the NIR appears
as stars, with no sign of hot dust because of normal $J-K$ colors,
we interpret it as an ongoing starburst, with a stellar
Red Super Giant (RSG) population.
This would imply that the ring feature is at least 10--15\,Myr old,
because RSGs onset at roughly that time \citep{sb99}.
However, the presence of hydrogen recombination lines
\citep[e.g.,][]{gonzalez96,hernandez03} means that star formation is
still ongoing, although the exact age is difficult to determine
(5-10\,Myr).

%The formation of star-forming rings is usually associated 
%with the link between bar-driven inflow and bar 
%resonances \citep[e.g.,][]{piner95}. 
%Also a simple oval feature (weak bar) may be able 
%to force the re-distribution of gas in galaxies
%\citep[e.g.,][]{buta96}.
%%  For NGC\,3147, classified as non-barred galaxy 
%%  in the optical, we found two inner $^{12}$CO rings with counterparts 
%%  both in the 8\,$\mu$m emission and in the FUV one \citep{vivi08}.
%%  Only near-infrared (NIR) observations reveal a weak bar just 
%%  contained inside the inner $^{12}$CO/8\,$\mu$m/FUV ring, which 
%%  may be the agent of the gas inflow in NGC\,3147.

In addition to the star-forming$+$stellar ring, the observations of
\citet{falcon06} suggest that the stars in the outer regions of the 
circumnuclear disk are counter-rotating relative to stars in the
inner disk and the ionized and molecular gas.
In this (putative) kinematically decoupled component (KDC),
the stars in the outer parts are redshifted toward the NE,
while the ionized$+$molecular gas and the stars in the inner parts
are redshifted toward the SW.
The separation between the two sets of kinematics
is at roughly 7-8\arcsec\ radius, so the ring could
be the separation between the two regimes.
The most likely explanation for this, as proposed by \citet{falcon06},
is that in \nnn\ we are witnessing the formation of a decoupled 
component as a result of an ongoing interaction; the gas in the inner 
parts could have been accreted from the companion
NGC\,5954.
The KDC could be in its early stages, as judged by our stellar
population age estimate of $\simgt$10-15\,Myr.
This age would be consistent with the onset of the interaction of
40\,Myr ago, as suggested by \citet{jenkins84};
this age from the time of closest approach is almost independent
of the details of the model.
This kind of counter-rotation has been also studied with simulations
\citep[e.g.,][]{paola08}, and the presence of the \hi\ tidal tail
gives some hint that the interaction in the pair NGC\,5953/5954
is prograde, i.e., that at least one of the disks involved
has its spin axis parallel to the orbital velocity of the merger
\citep{hibbard96,iono05}.

Alternatively, the observed counter-rotation could be only apparent 
and not real, due to a warped disk in the center.
The inner disk is not highly inclined, and a small tilt of the plane  
starting at a radius close to that of the ring, is able, because of 
projection effects, to reverse the apparent sense of rotation, 
without the presence of a true counter-rotation. 
A warped inner disk is expected from the close tidal 
interaction between NGC\,5953 with NGC\,5954, and the obvious tidal 
tails in optical and \hi.
 The warped disk hypothesis is supported by the fact that the ionised gas 
also, and not only the stars, shows some hint of counter-rotation. 
The existence of two gas flows counter-rotating in the same
disk, is a very unlikely situation, that could be at best transient.
The gas counter-rotation just outside the ring is visible in the 
Sauron velocity maps of \citet{falcon06}.
Indeed, the velocity maps in H$\beta$, and even more the [O{\sc iii}] lines 
reveal on the NE part of the major axis that the velocity from blueshifted  
turns to redshifted again, by an amount of $\sim$40 km\,s$^{-1}$.
The apparent counter-rotation might look more striking on the stellar
velocity maps, only because the stellar component is less 
rotation-dominated in the inner disk, because of the higher velocity dispersion. 
So the velocities on the major axis turn from -40 km\,s$^{-1}$ to 
40 km\,s$^{-1}$, while the gas is turning from -100 km\,s$^{-1}$
to 40 km\,s$^{-1}$. But it is clear that in the NE part of the
major axis, the stars and gas are co-rotating at the same velocities,
and apparently counter-rotate relative to the inner disk.
 Since the gas cannot be counter-rotating with itself for a long time, 
the interpretation of a warp in the outer disk is probably more realistic.
However, if the gas is actually counter-rotating at a radius of 1.6\,kpc, 
then it could arise from external accretion at that radius. 
This gas is configured in a full gaseous disk, of about 8\,kpc radius, 
as mapped in \hi\ by \citet{chengalur94}. 
We then can only witness a transient phase of settling of the accreted gas, 
which will soon align its rotation with the main disk, rotating in the same 
sense as the CO gas and stars in the center.

\begin{figure}
\includegraphics[angle=-90,width=0.4\textwidth]{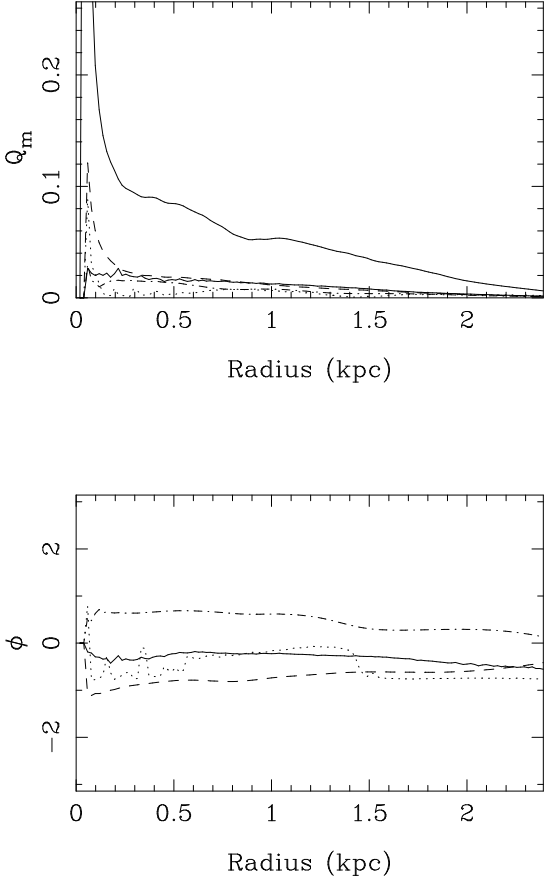}  
\caption{
The strength $Q$ (top) and phase $\Phi$ (bottom) of the $m$ = 1, 2, 3, 4 Fourier
components of the potential, derived from the NIR image.
The full lines correspond to $m=2$ and the total strength,
the dashed line to $m=1$, dot-dash to  $m=3$, and dots to  $m=4$.
}
\label{fig:potn5953}   
\end{figure} 

\subsection{Star formation in NGC\,5953}

Most of the star formation in \nnn\ occurs in the circumnuclear
region sampled by our $^{12}$CO observations.
The star-formation rate (SFR) given by H$\beta$  
within a FOV of 33\arcsec$\times$41\arcsec\
is 0.6\,M$_\odot$\,yr$^{-1}$ \citep{falcon06}. 
This corresponds to $\sim$0.024\,M$_\odot$\,yr$^{-1}$\,kpc$^{-2}$.
We measure a molecular mass of $1.7\times10^{9}$\,M$_\odot$
in a roughly similar region (see Sect. \ref{sec:comorphology}), 
which would give a molecular surface density of $\sim$60\,M$_\odot$\,pc$^{-2}$.
In a Kennicutt-Schmidt (KS) law diagram without the neutral atomic gas component,
this would place \nnn\ in the transition region between
normal spiral disks and circumnuclear starbursts 
\citep{kennicutt98}.
In fact, \hi\ observations show significant streaming motions over the 
inner 17\,kpc inconsistent with global rotation
\citep[][]{iono05,haan08}. Hence, the \hi\ probably does not take part 
in the star formation activity in this region.
It is probable that the enhanced star formation
in \nnn\ is caused by dynamical perturbations induced by the interaction.

In \nnn, the high H$_{2}$ surface density and the low SFR imply 
that the distribution of molecular gas (traced by CO) does not 
correlate locally with the SFR; 
the large reservoir of molecular gas is not converted into 
stars proportionally to the KS law on small spatial scales.
At present, the non-local correlation between SFR and gas is highly 
debated \citep{bigiel08,leroy08}. Other molecular transitions could be likely
better indicators of SFR than the total H$_{2}$ content traced by CO.
Tracers of dense molecular gas, such as HCN(1--0) and 
HCO$^{+}$(1--0) lines, are suspected to better correlate with the SFR
\citep[e.g.,][]{gao04a,gao04b,wu05,gracia-carpio08}.

\section{Gravitational torques on the molecular gas\label{sec:torques}}

In this section we study whether gravitational torques, 
derived from the stellar potential in the inner region of
\nnn,  can account for the gas kinematics
derived from CO and examine the efficiency of
gravitational torques exerted on the gas.
As described in previous NUGA papers \citep[e.g.,][]{santi05}, we
assume that NIR images give the best approximation for the total 
stellar mass distribution, being less affected than optical images 
by dust extinction or stellar population bias.

\begin{figure}
\includegraphics[angle=-90,width=0.45\textwidth]{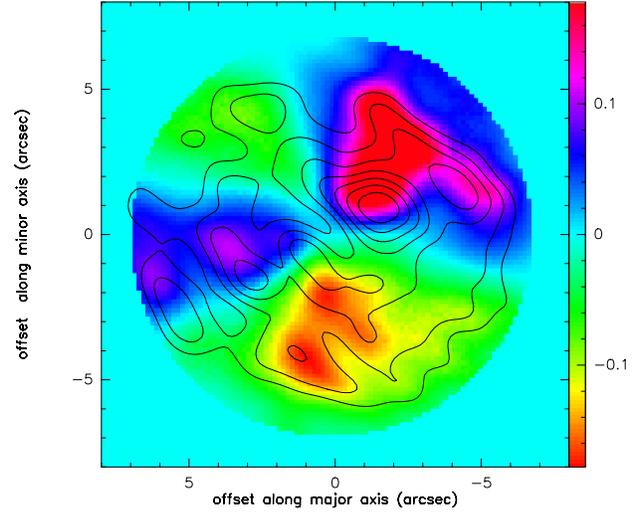}
\caption{
The $^{12}$CO(1--0) contours are overlaid onto the gravitational torque
map (t(x,y)~$\times$~$\Sigma$(x,y), as defined in text) in the center of NGC~5953.
The torque map (color scale) is plotted on a symmetric palette (wedge)
although the maximum positive torque is 50\% higher in absolute value
then the negative torque.
%The derived torques change sign as expected in a {\it butterfly} diagram,
%delineating four quadrants. There is more molecular mass in the
%positive quadrants, and the resulting net torque is positive.
The map is deprojected, and rotated so that 
the major axis of the galaxy is oriented parallel to the abscissa Ox.
}
\label{fig:n59-torq1}
\end{figure}

\begin{figure}[ht]
\includegraphics[angle=-90,width=0.45\textwidth]{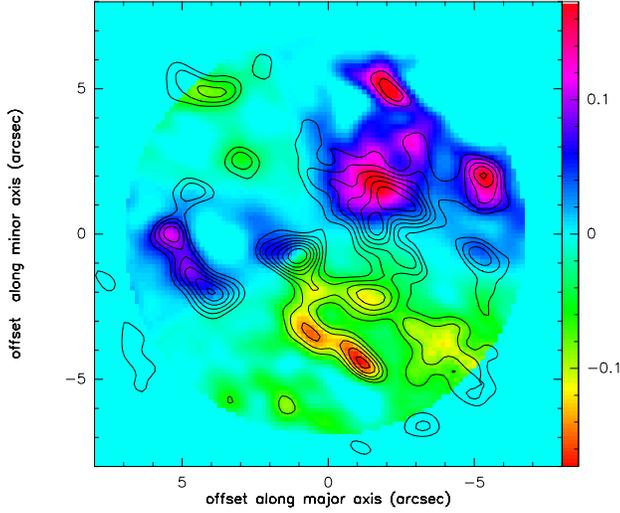}
\caption{
Same as Fig. \ref{fig:n59-torq1} for the $^{12}$CO(2--1)
emission taken as tracer of gas surface density. 
}
\label{fig:n59-torq2}
\end{figure}

\subsection{Evaluation of the gravitational potential\label{sec:grav-pot}}

We computed the torques using \spitzer/IRAC and 
ARNICA images.  They both yield comparable
results, but we use in the following only the NOT/ARNICA images
which give more spatial resolution.
 We perform the subtraction of foreground stars, 
deprojection, and resampling,  as described in
other NUGA papers \citep[e.g.,][]{santi05}.

We repeat here some definitions and assumptions used to evaluate 
the gravitational torques.
The NIR image is completed in the vertical 
dimension by assuming an isothermal plane model with a constant 
scale height, equal to $\sim$1/12th of the radial 
scale-length of the image. 
The potential is derived by a Fourier transform method 
and assuming a constant mass-to-light (M/L) ratio whose 
value is chosen to better reproduce the observed $^{12}$CO 
rotation curve.
Beyond a radius of 20\arcsec (or 5.4 kpc in diameter), the mass density is set to 
0, thus suppressing any spurious $m=4$ terms. This assumption
is sufficient to compute the potential over the 
PdBI $^{12}$CO(1--0) primary beam.

\begin{figure*}
\includegraphics[width=\textwidth,bb=15 130 580 300]{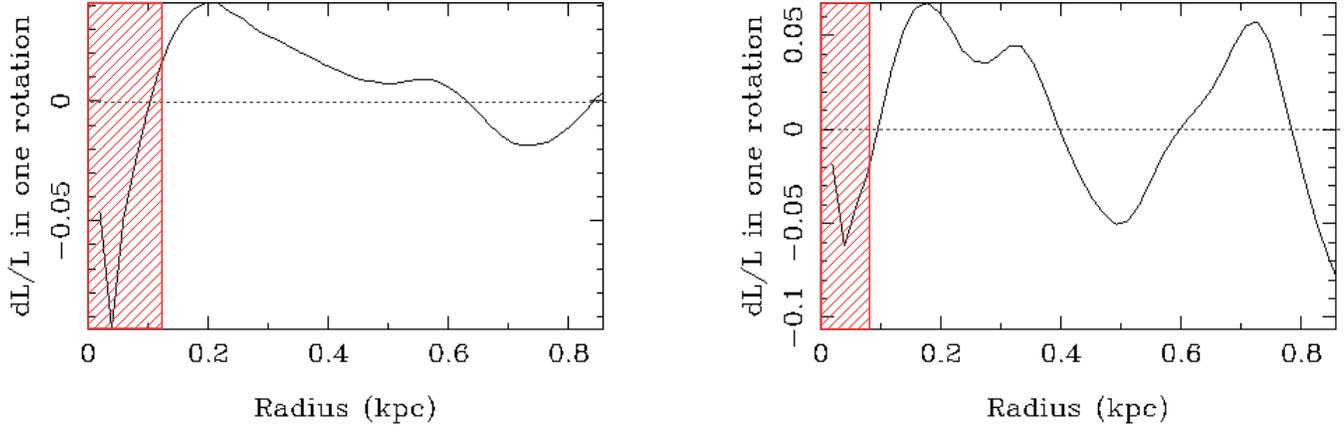}
\caption{
The torque, or more precisely the fraction of the 
angular momentum transferred from/to the gas in one rotation 
--$dL/L$-- is plotted for $^{12}$CO(1--0) (left) and $^{12}$CO(2--1) 
(right). The (red) dashed area corresponds to the resolution limit of our 
observations.
In the left panel, the CO(1--0) map resolution is the limiting factor, 
while, on the right, the CO(2--1) and the similar NIR image resolution are the constraining factors.}
\label{fig:gastor5953}   
\end{figure*} 

To characterize the non-axisymmetry of the potential,
we  expand its $\Phi(R,\theta)$ in Fourier 
components ($m$-modes), following \citet{francoise81}:
$$
\Phi(R,\theta) = \Phi_0(R) + \sum_{m=1}^\infty \Phi_m(R) \cos [m \theta - \phi_m(R)]
$$
\noindent
where $\Phi_m(R)$ and $\phi_m(R)$ are the amplitude and phase of the 
$m$-mode, respectively.

The strength of each $m$-Fourier component, $Q_m(R)$, is defined 
by the ratio between tangential and radial forces, 
$Q_m(R)=m \Phi_m / R | F_0(R) |$.
The strength of the total non-axisymmetric perturbation is defined by:
$$
Q_T(R) = {F_T^{max}(R) \over F_0(R)} 
$$
\noindent
where $F_T^{max}(R)$ and $F_{0}(R)$ represent the maximum 
amplitude of the tangential force and the mean axisymmetric 
radial force, respectively.
Figures \ref{fig:potn5953} show the  strengths and phases vs. 
radius for the first $m$ components. There is only a weak
$m=2$ perturbation; the strongest one is $m=1$.

\subsection{Evaluation of gravity torques\label{sec:grav-tor}}
From the derivatives of $\Phi(R,\theta)$ on each pixel we obtain  
the forces per unit mass ($F_x$ and $F_y$) and therefore
the torques per unit mass $t(x,y)$ can be computed by:
$$
t(x,y) = x~F_y -y~F_x.
$$
The torque map is oriented according to the sense of rotation 
in the plane of the galaxy, and the combination of the torque 
map and the gas density $\Sigma$ map allows us to derive 
the net effect on the gas at each radius.
The gravitational torque map weighted by the gas surface 
density $t(x,y)\times \Sigma(x,y)$, normalized to its maximum 
value, is shown in Figures \ref{fig:n59-torq1} and \ref{fig:n59-torq2} 
for $^{12}$CO(1--0) and $^{12}$CO(2--1), respectively.
The observed gas distribution is representative of the time 
spent by a molecular cloud on a typical orbit at this location. 

To estimate the efficiency of the radial gas flow induced by the torques, 
we first compute the torque per unit mass averaged over azimuth, 
using $\Sigma(x,y)$ as the actual weighting function:
$$
t(R) = \frac{\int_\theta \Sigma(x,y)\times(x~F_y -y~F_x)}{\int_\theta \Sigma(x,y)}
$$
where $t(R)$ is, for definition, the time derivative of the specific 
angular momentum $L$ of the gas averaged azimuthally, 
$t(R)$=$dL/dt~\vert_\theta$.   
Now to have dimensionless  quantities, we normalize this variation 
of angular momentum per unit time to the angular momentum at 
this radius and to the rotation period.
We then estimate the efficiency of the gas flow as the average 
fraction of the gas specific angular momentum transferred 
in one rotation ($T_{rot}$) by the stellar potential, as a 
function of radius:
$$
{\Delta L\over L}=\left.{dL\over dt}~\right\vert_\theta\times \left.{1\over L}~\right\vert_\theta\times 
T_{rot}={t(R)\over L_\theta}\times T_{rot}
$$
\noindent
where $L_\theta$ is assumed to be well represented by its axisymmetric 
estimate, $L_\theta=R\times v_{rot}$.
%Figures \ref{fig:gastor5953} show $\Delta L/L$ curves for NGC\,5953 
%derived from the $^{12}$CO(1--0) (left) and the $^{12}$CO(2--1) 
%(right) data. 
%This figure shows that the torques are predominantly positive in both  
%$^{12}$CO(1--0) and $^{12}$CO(2--1).
%Nothing can be said for the negative torques  
%inside a radius of $\sim$0.1\,kpc, below our 
%resolution.
%In $^{12}$CO(2--1) the torques are noisier than 
%in $^{12}$CO(1--0), maybe due to a more clumpy emission. 
Figures \ref{fig:gastor5953} show $\Delta L/L$ curves for NGC\,5953 
derived from the $^{12}$CO(1--0) (left) and the $^{12}$CO(2--1) 
(right) data. 
The (red) dashed area corresponds to the resolution limit of our 
observations. In the left panel, the CO(1--0) map resolution ($\sim$120\,pc) 
is the limiting factor, while, on the right, the CO(2--1) map resolution ($\sim$60\,pc) and the similar NIR image resolution ($\sim$70\,pc) are the constraining factors.
This figure shows that the torques are weak and predominantly positive 
between $\sim$100--400\,pc in both $^{12}$CO(1--0) and $^{12}$CO(2--1).
Nothing can be said for the negative torques  
inside a radius of $\sim$120\,pc for $^{12}$CO(1--0) and 
$\sim$70\, pc, below our resolution.
In $^{12}$CO(2--1) the torques are noisier than 
in $^{12}$CO(1--0), maybe because of a more clumpy emission.

%The gravitational torque maps, displayed in Figs. \ref{fig:n59-torq1} and
%\ref{fig:n59-torq2}, show that the derived torques change sign following 
%a characteristic 2D {\it butterfly} pattern. 

In summary, torques are predominantly positive in the region 
$\sim$100--400\,pc for both lines, and the gas is not apparently fueling 
the central $\sim$100\,pc region down to the effective spatial resolution 
of our observations.
%In summary, predominantly positive torques indicate that the gas is 
%not flowing into the center of \nnn\ and surely not fueling the AGN. 
In addition, the absolute value of the torques is very small,
less than 5\% of the gas angular momentum is exchanged in each 
rotation. This is due to the very regular and almost axisymmetric 
total mass and gas distributions in the center of the galaxy.

\section{Summary and conclusions \label{sec:conclusions}}
The molecular gas in the Seyfert 2/LINER galaxy \nnn\ has been 
mapped with high resolution (2\farcs1 $\times$ 1\farcs4 
for the $^{12}$CO(1--0) line and 1\farcs1 $\times$ 0\farcs7
for the $^{12}$CO(2--1) line) inside a radius of 
$\sim$20\arcsec\ ($\sim$2.7\,kpc).
The $^{12}$CO emission is distributed over a disk 
of diameter of $\sim$16$^{\prime\prime}$ with several peaks 
randomly distributed. The strongest intensity peak is not located in 
the center of the galaxy but shifted $\sim$2-3$^{\prime\prime}$ 
toward the west/southwest.
The kinematics of the molecular gas show a general 
regularity with some local wiggles especially to the west of 
the nucleus maybe associated with the intensity peak toward 
the SW.

By comparing the molecular gas distribution with 
observations at other wavelengths, we studied correlations
between different tracers of the ISM.
Optical and NIR morphology has been studied by analyzing 
different images. The F606W/HST  broad-band optical image of 
NGC\,5953 shows  a flocculent spiral structure and  an ``S'' 
like-bar of $\sim$250\,pc and $\simgt$60\,pc in size, respectively.  
These structures are absent in the $1.6\,{\rm \mu m}$
(\hst/NICMOS/F160W) image and in the IRAC $8\,{\rm \mu m}$ 
``dust-only'' one; instead in the optical and at 
$1.6\,{\rm \mu m}$ we identified at $\sim$3\arcsec\ west 
of center the foreground star studied by \citet{piraf90}.
The stellar structure has been investigated by performing
a two-dimensional B/D decomposition on the ground-based 
$H$-band image and IRAC 3.6\,$\mu$m image.
Both sets of residuals show a circumnuclear stellar ring with a radius of 
$\sim$5-7\arcsec, approximately coincident in size with 
the $^{12}$CO disk.
The presence of the ring in both sets of residuals suggests that it
is not the typical ``doughnut'' associated with the diffraction 
limit and IRAC PRF incompatibilities, but rather is a real feature.
The size of this stellar ring is the same of the circumnuclear 
star-forming ring seen in H$\alpha$ with previous 
observations. 
In addition, since this ring is visible in the NIR as stars 
but not as hot dust emission ($J-K$), we interpret it as
a RSG population at least 10\,Myr old.

There is an apparent counter-rotation between gas and stars
inside the ring, and stars outside.
The ring could be the separation between the two kinematically 
decoupled components (KDC). 
The formation of a KDC could be explained 
as a result of the ongoing interaction, and could be in its 
early stages. Alternatively, this apparent counter-rotation
could be due to a warp of the plane of the disk.
The ring would then be the start of the warp.

%Using NIR images we found that gravity torques acting on the gas are 
%predominantly positive in both  $^{12}$CO(1--0) and $^{12}$CO(2--1), 
%indicating that the gas is not flowing into the center of \nnn.
%In addition, the absolute values of these torques are very
%small, certainly due to the almost axisymmetric total mass
%and gas distributions in the center of the galaxy.
%Despite the extremely high molecular gas fraction ($\sim$22$\%$) 
%and the consequently large gas reservoir, the AGN in \nnn\ is apparently 
%not being actively fueled in the current epoch.

Using NIR images we found that gravity torques acting on the gas are 
predominantly positive in the region 
$\sim$100--400\,pc in both  $^{12}$CO(1--0) and $^{12}$CO(2--1), 
indicating that the gas is not fueling the central 
$\sim$100\,pc region, down to the effective spatial resolution of our 
observations.
In addition, the absolute values of these torques are very
small, certainly due to the almost axisymmetric total mass
and gas distributions in the center of the galaxy.
The AGN in \nnn\ is apparently 
not being actively fueled in the current epoch.

\begin{acknowledgements}
The authors would like to thank the anonymous referee,
whose comments have been useful for improving the original
version of the paper.
We thank the scientific and technical staff at IRAM for their 
work in making our 30\,m and PdBI observations possible.
V. Casasola wishes to thank Riccardo Cesaroni and Claudio Codella for 
help provided during the data reduction.
This research has made use of the NASA/IPAC Extragalactic Database (NED),
HyperLeda Database, {\it IRAS} Catalog, \spitzer\ archive, and
Hubble Legacy Archive.
\end{acknowledgements}

%\clearpage


\begin{thebibliography}{}
\bibitem[Bigiel et al.(2008)]{bigiel08} Bigiel, F., Leroy, A., 
Walter, F., Brinks, E., de Blok, W.~J.~G., Madore, B., 
\& Thornley, M.~D.\ 2008, \aj, 136, 2846 

\bibitem[Boone et 
al.(2007)]{fred07} Boone, F., et al.\ 2007, \aap, 471, 113 (NUGA VII)

\bibitem[Boselli et al.(2002)]{boselli02} Boselli, A., Lequeux, J., \& Gavazzi, G.\ 2002, \apss, 281, 127 

\bibitem[Braine \& Combes(1992)]{braine92} Braine, J., \& 
Combes, F.\ 1992, \aap, 264, 433

\bibitem[Cardelli et al.(1989)]{cardelli89} Cardelli, J.~A., 
Clayton, G.~C., \& Mathis, J.~S.\ 1989, \apj, 345, 245 

\bibitem[Casasola et 
al.(2004)]{vivi04} Casasola, V., Bettoni, D., \& Galletta, G.\ 2004, \aap, 422, 941

\bibitem[Casasola et 
al.(2007)]{vivi07} Casasola, V., Combes, F., Bettoni, D., \& Galletta, G.\ 2007, \aap, 473, 771 

\bibitem[Casasola et al.(2008)]{vivi08} Casasola, V., Combes, F,. Garc{\'{\i}}a-Burillo, S.,
Hunt, L.~K., L\'eon, S., \&  Baker, A. J., \ 2008, \aap, 490, 61

\bibitem[Chengalur et al.(1994)]{chengalur94} Chengalur, J.~N., 
Salpeter, E.~E., \& Terzian, Y.\ 1994, \aj, 107, 1984

\bibitem[Combes \& Sanders(1981)]{francoise81} Combes, F., \& 
Sanders, R.~H.\ 1981, \aap, 96, 164 

\bibitem[Combes \& Gerin(1985)]{francoise85} Combes, F., \& Gerin, 
M.\ 1985, \aap, 150, 327 

\bibitem[Combes(2001)]{francoise01} Combes, F.\ 2001, Advanced 
Lectures on the Starburst-AGN , 223 

\bibitem[Combes et al.(2004)]{francoise04} 
Combes, F., et al.\ 2004, \aap, 414, 857 (NUGA II)

\bibitem[Combes et al.(2009)]{francoise09} Combes, F., et al.\ 2009, \aap, 
503, 73 (NUGA XII)

\bibitem[de Jong(1996)]{dejong96} de Jong, R.~S.\ 1996, \aap, 313, 377 

\bibitem[di Matteo et al.(2008)]{paola08} di Matteo, P., Combes, F., 
Melchior, A.-L., \& Semelin, B.\ 2008, \aap, 477, 437 

\bibitem[Falc{\'o}n-Barroso et al.(2006)]{falcon06} 
Falc{\'o}n-Barroso, J., et al.\ 2006, \mnras, 369, 529 

\bibitem[Ferrarese et al.(2001)]{ferrarese01} Ferrarese, L., Pogge, 
R.~W., Peterson, B.~M., Merritt, D., Wandel, A., \& Joseph, C.~L.\ 2001, 
\apjl, 555, L79 

\bibitem[Friedli \& Martinet(1993)]{friedli93} Friedli, D., \& 
Martinet, L.\ 1993, \aap, 277, 27 

\bibitem[Gao \& Solomon(2004a)]{gao04a} Gao, Y., \& Solomon, 
P.~M.\ 2004a, \apj, 606, 271 

\bibitem[Gao \& Solomon(2004b)]{gao04b} Gao, Y., \& Solomon, P.~M.\ 2004b, 
\apjs, 152, 63 

\bibitem[Garc{\'{\i}}a-Burillo et al.(1993)]{santi93} Garcia-Burillo, S., Guelin, M., \& Cernicharo, J.\ 1993, \aap, 274, 123 

\bibitem[Garc{\'{\i}}a-Burillo et al.(2000)]{santi00} 
Garc{\'{\i}}a-Burillo, S., Sempere, M.~J., Combes, F., Hunt, L.~K., \& 
Neri, R.\ 2000, \aap, 363, 869 

\bibitem[Garc{\'{\i}}a-Burillo et al.(2003)]{santi03} 
Garc{\'{\i}}a-Burillo, S., et al.\ 2003, \aap, 407, 485 (NUGA I)

\bibitem[Garc{\'{\i}}a-Burillo et al.(2005)]{santi05} 
Garc{\'{\i}}a-Burillo, S., Combes, F., Schinnerer, E., Boone, F., \& Hunt, 
L.~K.\ 2005, \aap, 441, 1011 (NUGA IV)

\bibitem[Garc{\'{\i}}a-Burillo et al.(2009)]{santi09} 
Garc{\'{\i}}a-Burillo, S., et al.\ 2009, \aap, 496, 85 (NUGA XII)

\bibitem[Graci{\'a}-Carpio et al.(2008)]{gracia-carpio08} Graci{\'a}-Carpio, J., Garc{\'{\i}}a-Burillo, S., Planesas, P., Fuente, A., \& Usero, A.\ 2008, \aap, 479, 703 

\bibitem[Gonzalez Delgado 
\& Perez(1996)]{gonzalez96} Gonzalez Delgado, R.~M., \& Perez, E.\ 1996, \mnras, 
281, 781

\bibitem[Guilloteau \& Lucas(2000)]{guilloteau} Guilloteau, S., \&
Lucas, R.\ 2000, Imaging at Radio through Submillimeter Wavelengths, 217,
299

\bibitem[Haan et al.(2007)]{haan07} Haan, S., Schinnerer, E., 
Mundell, C.~G., Combes, F., Garcia-Burillo, S., 
\& Emsellem, E.\ 2007, Astronomische Nachrichten, 328, 675 

\bibitem[Haan et al.(2008)]{haan08} Haan, S., Schinnerer, E., 
Mundell, C.~G., Garc{\'{\i}}a-Burillo, S., 
\& Combes, F.\ 2008, \aj, 135, 232 

\bibitem[Heckman et al.(1986)]{heckman86} Heckman, T.~M., Smith, 
E.~P., Baum, S.~A., et al.: 1986, \apj, 311, 526 

\bibitem[Heckman et al.(2004)]{heckman04} Heckman, T.~M., 
Kauffmann, G., Brinchmann, J., Charlot, S., Tremonti, C., \& White, 
S.~D.~M.\ 2004, \apj, 613, 109 

\bibitem[Helou et al.(2004)]{helou04} Helou, G., et al.\ 2004, 
\apjs, 154, 253 

\bibitem[Hern{\'a}ndez-Toledo et 
al.(2003)]{hernandez03} Hern{\'a}ndez-Toledo, H.~M., Fuentes-Carrera, I., 
Rosado, M., Cruz-Gonz{\'a}lez, I., Franco-Balderas, A., \& Dultzin-Hacyan, 
D.\ 2003, \aap, 412, 669 

\bibitem[Hibbard \& van Gorkom(1996)]{hibbard96} 
Hibbard, J.~E., \& van Gorkom, J.~H.\ 1996, \aj, 111, 655 

\bibitem[Hopkins \& Hernquist(2006)]{hopkins06} Hopkins, P.~F., 
\& Hernquist, L.\ 2006, \apjs, 166, 1 

\bibitem[Hunt et al.(1999)]{leslie99b} Hunt, L.~K., Malkan, 
M.~A., Rush, B., Bicay, M.~D., Nelson, B.~O., Stanga, R.~M., 
\& Webb, W.\ 1999, \apjs, 125, 349 

\bibitem[Hunt 
\& Malkan(2004)]{leslie04b} Hunt, L.~K., \& Malkan, M.~A.\ 2004, \apj, 616, 707 

\bibitem[Hunt et 
al.(2004)]{leslie04a} Hunt, L.~K., Pierini, D., \& Giovanardi, C.\ 2004, \aap, 414, 905 

\bibitem[Hunt et al.(2008)]{leslie08} Hunt, L.~K., et al.\ 2008, \aap, 482, 133 
(NUGA IX)

\bibitem[Iono et al.(2005)]{iono05} Iono, D., Yun, M.~S., 
\& Ho, P.~T.~P.\ 2005, \apjs, 158, 1

\bibitem[Janiuk et al.(2004)]{janiuk04} Janiuk, A., Czerny, B., 
Siemiginowska, A., \& Szczerba, R.\ 2004, \apj, 602, 595 

\bibitem[Jenkins(1984)]{jenkins84} Jenkins, C.~R.\ 1984, \apj, 277, 501 

\bibitem[Jogee et al.(2005)]{jogee05} Jogee, S., Scoville, N., 
\& Kenney, J.~D.~P.\ 2005, \apj, 630, 837 

\bibitem[Jogee(2006)]{jogee06} Jogee, S.\ 2006, Physics of 
Active Galactic Nuclei at all Scales, 693, 143, (astro-ph/0408383) 

\bibitem[Kenney 
\& Young(1986)]{Kenney86} Kenney, J.~D., \& Young, J.~S.\ 1986, \apjl, 301, L13

\bibitem[Kennicutt(1998)]{kennicutt98} Kennicutt, R.~C., Jr.\ 1998, \araa, 36, 189

\bibitem[King 
\& Pringle(2007)]{king07} King, A.~R., \& Pringle, J.~E.\ 2007, \mnras, 377, L25 

\bibitem[Knapen et al.(2002)]{knapen02} Knapen, J.~H., 
P{\'e}rez-Ram{\'{\i}}rez, D., \& Laine, S.\ 2002, \mnras, 337, 808 

\bibitem[Kohno et al.(2003)]{kohno03} 
Kohno, K., Ishizuki, S., Matsushita, S., Vila-Vilaro, B., Kawabe, R.: 2003,
PASJ   55, L1

\bibitem[Krips et al.(2005)]{melanie05} 
Krips, M., et al.\ 2005, \aap, 442, 479 (NUGA III)

\bibitem[Krips et al.(2007a)]{melanie07a} Krips, M., et al.\ 2007a, \aap, 464, 553 

\bibitem[Krips et al.(2007b)]{melanie07b} Krips, M., Neri R., Garc{\'{\i}}a-Burillo, S.,: 
2007b, A\&A 468, L63, (NUGA VI)   % HCN in N6951

%\bibitem[Laurikainen et al.(2004)]{laurikainen04} Laurikainen, E., 
%Salo, H., \& Buta, R.\ 2004, \apj, 607, 103 

\bibitem[Leitherer et al.(1999)]{sb99} Leitherer, C., et 
al.\ 1999, \apjs, 123, 3 

\bibitem[Leroy et al.(2008)]{leroy08} Leroy, A.~K., Walter, F., 
Brinks, E., Bigiel, F., de Blok, W.~J.~G., Madore, B., 
\& Thornley, M.~D.\ 2008, \aj, 136, 2782 

\bibitem[Makovoz \& Marleau(2005)]{makovoz05} Makovoz, D., \& 
Marleau, F.~R.\ 2005, \pasp, 117, 1113 

\bibitem[Malkan et al.(1998)]{malkan98} Malkan, M.~A., Gorjian, 
V., \& Tam, R.\ 1998, \apjs, 117, 25 

\bibitem[Maloney \& Black(1988)]{maloney88} Maloney, P., 
\& Black, J.~H.\ 1988, \apj, 325, 389 

\bibitem[Marecki et al.(2003)]{marecki03} Marecki, A., Spencer, 
R.~E., \& Kunert, M.\ 2003, Publications of the Astronomical Society of 
Australia, 20, 46 

\bibitem[Moriondo et al.(1998)]{moriondo98} 
Moriondo, G., Giovanardi, C., \& Hunt, L.~K.\ 1998, \aaps, 130, 81 

\bibitem[Mulchaey \& Regan(1997)]{mulchaey97} Mulchaey, J.~S., \& 
Regan, M.~W.\ 1997, \apjl, 482, L135 

\bibitem[Nakai 
\& Kuno(1995)]{nakai95} Nakai, N., \& Kuno, N.\ 1995, \pasj, 47, 761 

\bibitem[Narayanan et al.(2006)]{narayanan06} Narayanan, D., et 
al.\ 2006, \apjl, 642, L107 

% \bibitem[Pahre et al.(2004)]{pahre04} Pahre, M.~A., Ashby, 
% M.~L.~N., Fazio, G.~G., \& Willner, S.~P.\ 2004, \apjs, 154, 235 

\bibitem[Paturel et 
al.(2003)]{paturel03} Paturel, G., Petit, C., Prugniel, P., 
Theureau, G., Rousseau, J., Brouty, M., Dubois, P., 
\& Cambr{\'e}sy, L.\ 2003, \aap, 412, 45

\bibitem[Peng et al.(2002)]{peng02} Peng, C.~Y., Ho, L.~C., 
Impey, C.~D., \& Rix, H.-W.\ 2002, \aj, 124, 266 

\bibitem[Rafanelli et al.(1990)]{piraf90} Rafanelli, P., 
Osterbrock, D.~E., \& Pogge, R.~W.\ 1990, \aj, 99, 53 

\bibitem[Rampazzo et 
al.(1995)]{rampazzo95} Rampazzo, R., Reduzzi, L., 
Sulentic, J.~W., \& Madejsky, R.\ 1995, \aaps, 110, 131 

\bibitem[Regan 
\& Mulchaey(1999)]{regan99} Regan, M.~W., \& Mulchaey, J.~S.\ 1999, \aj, 117, 2676 

\bibitem[Sakamoto et al.(1999)]{sakamoto99} Sakamoto, K., Okumura, 
S.~K., Ishizuki, S., \& Scoville, N.~Z.\ 1999, \apj, 525, 691 

\bibitem[Schinnerer et al.(2000a)]{schinnerer00a} Schinnerer, E., 
Eckart, A., \& Tacconi, L.~J.\ 2000a, \apj, 533, 826 

\bibitem[Schinnerer et al.(2000b)]{schinnerer00b} Schinnerer, E., 
Eckart, A., Tacconi, L.~J., Genzel, R., \& Downes, D.\ 2000b, \apj, 533, 850 

\bibitem[Shu et al.(1990)]{shu90} Shu, F.~H., Tremaine, S., 
Adams, F.~C., \& Ruden, S.~P.\ 1990, \apj, 358, 495 

\bibitem[Solomon \& Barrett(1991)]{solomon91} Solomon, P.~M., \&
Barrett, J.~W.\ 1991, Dynamics of Galaxies and Their Molecular Cloud
Distributions, 146, 235

\bibitem[Veilleux et al.(1995)]{veilleux95} Veilleux, S., Kim, 
D.-C., Sanders, D.~B., Mazzarella, J.~M., 
\& Soifer, B.~T.\ 1995, \apjs, 98, 171

\bibitem[Vollmer et 
al.(2001)]{Vollmer01} Vollmer, B., Braine, J., Balkowski, C., Cayatte, V., \& Duschl, W.~J.\ 2001, \aap, 374, 824 

\bibitem[Wilson(1995)]{wilson95} Wilson, C.~D.\ 1995, \apjl, 
448, L97 

\bibitem[Wu et al.(2005)]{wu05} Wu, J., Evans, N.~J., II, 
Gao, Y., Solomon, P.~M., Shirley, Y.~L., 
\& Vanden Bout, P.~A.\ 2005, \apjl, 635, L173 

\bibitem[Yao et al.(2003)]{yao03} Yao, L., Seaquist, E.~R., 
Kuno, N., \& Dunne, L.\ 2003, \apj, 588, 771 

\bibitem[Young et al.(1995)]{young95} Young, J.~S., et al.\ 1995, \apjs, 98, 219 

\bibitem[Zhu et al.(1999)]{zhu99} Zhu, M., Seaquist, E.~R., 
Davoust, E., Frayer, D.~T., \& Bushouse, H.~A.\ 1999, \aj, 118, 145 


\end{thebibliography}
\end{document}